# Recent methods from statistical inference and machine learning to improve integrative modeling of macromolecular assemblies

Shreyas Arvindekar[1,*], Kartik Majila[1,*], Shruthi Viswanath[1,#]

[1]National Center for Biological Sciences, Tata Institute of Fundamental Research, Bangalore, India 560065.

[*]Contributed equally. Author order was determined by a coin toss.

[#]Corresponding author. Email: shruthiv@ncbs.res.in

# List of symbols and abbreviations

1. PDB: Protein Data Bank
2. NMR: Nuclear Magnetic Resonance
3. Cryo-EM: Cryo-Electron Microscopy
4. Cryo-ET: Cryo-Electron Tomography
5. EM: Electron Microscopy
6. SAXS: Small Angle X-ray Scattering
7. FRET: Forster Resonance Energy Transfer
8. XLMS: Chemical crosslinking by mass spectrometry
9. Y2H: Yeast Two-Hybrid
10. co-IP: co-immunoprecipitation
11. IMP: Integrative Modeling Platform
12. AF2: Alphafold2
13. HADDOCK: High Ambiguity Driven biomolecular DOCKing
14. PDB-Dev: Protein Data Bank-development
15. wwPDB: worldwide Protein Data Bank
16. RMSD: Root Mean Square Deviation
17. iRMSD: interface Root Mean Square Deviation
18. LPD: Localization Probability Density
19. HDBSCAN: Hierarchical Database Scan
20. HDXMS: Hydrogen Deuterium eXchange-Mass Spectrometry
21. MSA: Multiple Sequence Alignment
22. IDP: Intrinsically Disordered Protein
23. MD: Molecular Dynamics
24. AIR: Ambiguous Interaction Restraint
25. MDFF: Molecular Dynamics Flexible Fitting
26. TEMpyREFF: Template and Electron Microscopy comparison using Python Responsibility-based flexible fitting
27. DeepMainMAST: Deep Main-chain Model trAcing from Spanning Tree
28. MainMAST:Main-chain Model trAcing from Spanning Tree
29. ReMDFF: resolution-exchange molecular dynamics flexible fitting
30. MELD: modeling employing limited data
31. Emap2sec: EM map to secondary structure

32. PLUMED-ISDB: PLUMED Integrative Structural and Dynamical Biology
33. EROS: Ensemble Refinement for SAXS
34. BioEN: Bayesian inference of ENsembles
35. SCAP-XL: Serial Capture Affinity Purification along with XLMS
36. SEC-MALS: Size Exclusion Chromatography Multi-Angle Light Scattering
37. DIA-MS: Data Independent Acquisition Mass Spectrometry
38. SMC: Structural Maintenance of Chromosomes
39. NuRD: Nucleosome Remodeling and Deacetylase
40. NuDe: Nucleosome Deacetylase
41. DMT: Doublet Microtubules
42. SMT: Singlet Microtubules
43. MIP: Microtubule Inner Proteins
44. NPC: Nuclear Pore Complex
45. CR: Cytoplasmic Ring
46. CFNC: Cytoplasmic Filament NuCleoporin
47. FCS: Fluorescence Correlation Spectroscopy
48. T3SS: Type III Secretion System
49. Nbs: Nanobodies
50. RBD: receptor-binding domain
51. hcAb: heavy chain-only functional antibodies
52. TCR: T-cell receptor
53. CNN: Convolutional Neural Networks
54. VAE: Variational Auto Encoder
55. cryoDRGN: cryo-Deep Reconstructing Generative Networks
56. MNXL: Matched and Non-accessible Crosslink
57. BiCEPs: Bayesian Inference of Conformational Populations
58. StOP: Stochastic Optimization of Parameters
59. PrISM: PRecision for Integrative Structural Models
60. NestOR: Nested Sampling for Optimizing Representations
61. RF: RosettaFold
62. AFDB: AlphaFold Database
63. i-pTM: interface-predicted Template Match
64. X-EISD: extended Experimental Inferential Structure Determination
65. DynamICE: dynamic IDP creator with experimental restraints

66. GRNN: generative recurrent neural network
67. RL: reinforcement learning
68. RFDiffusion: RosettaFold Diffusion
69. STA: SubTomogram Averaging
70. SHREC: Shape Retrieval Contest
71. DeePICT: Deep Picker in Context
72. YOLO: You Only Look Once
73. ProtoNet: Prototypical Network
74. Protonet-CE: Protonet with Combined Embedding
75. MPP: Multi-Pattern Pursuit
76. DoG: Difference-of-Gaussian
77. DISCA: Deep Iterative Subtomogram Clustering Approach
78. EM: Expectation Maximization

# Abstract

Integrative modeling of macromolecular assemblies allows for structural characterization of large assemblies that are recalcitrant to direct experimental observation. A Bayesian inference approach facilitates combining data from complementary experiments along with physical principles, statistics of known structures, and prior models, for structure determination. Here, we review recent methods for integrative modeling based on statistical inference and machine learning. These methods improve over the current state-of-the-art by enhancing the data collection, optimizing coarse-grained model representations, making scoring functions more accurate, sampling more efficient, and model analysis more rigorous. We also discuss three new frontiers in integrative modeling: incorporating recent deep learning-based methods, integrative modeling with *in situ* data, and metamodeling.

# Introduction to Integrative Structure Modeling

Structural biology was conceived in the 1950s [1]. It took approximately twenty years for the first protein structure to be solved – Myoglobin (1958) [1–3]. Due to advancements in experimental and computational methodologies, structural biology has expanded to embrace structures at multiple scales and multiple states [2]. The number of structures deposited in the Protein Data Bank (PDB) has grown exponentially from 684 in 1991 to 211,103 in 2023 [2, 4] (https://www.rcsb.org/).

Integrative structural modeling aims to determine the structure of biomolecular systems by integrating information from multiple sources [1, 5–9]. This may entail experimental data, physical principles, statistics of previous structures and sequences, and prior models [1, 10]. The data used for integrative modeling could be of varying resolutions, sparsity, and noise, and could result from multiple structural states. However, different sources of data may provide complementary information about the orientation and/or localization of the subunits of the system [1]. This approach in principle maximizes the accuracy, precision, completeness, and efficiency of structure determination [1, 9, 10].

Any single structure determination technique may not be sufficient for determining the structure at high resolution, especially in the case of large macromolecular assemblies, and dynamic, or intrinsically disordered proteins. X-ray crystallography captures only static structures, requires the samples in large amounts, and requires them to be crystallizable [11]. Nuclear magnetic resonance (NMR) can capture the dynamics of structure but suffers from on the size of the system. Cryo-electron microscopy (cryo-EM) and small angle X-ray scattering (SAXS) can be limited by low resolution [5]. Techniques like cross-linking mass spectrometry (XLMS), yeast two-hybrid (Y2H), co-immunoprecipitation (co-IP), and Förster resonance energy transfer (FRET) provide protein-protein interaction data. Even computational methods suffer from poor accuracy of prediction due to reasons such as lack of homologous templates (homology modeling) and poor-quality sequence alignments (co-evolutionary information-based methods like AlphaFold2). By integrating information from multiple such techniques, one can overcome the limitations for each of these individually and allow for determining structures at high resolution, accuracy, and coverage [12]. Here, we describe integrative methods for static and dynamical modeling of macromolecular systems.

## Static structure modeling

Biological systems are dynamic by nature; however, static structural models play a crucial role in understanding the structure and mechanisms of various biological systems and processes. Here, we describe integrative modeling methods for modeling static structures of biological systems.

### Integrative Modeling Platform (IMP)

IMP, uses a Bayesian inference approach for modeling biomolecular systems, ranging from small peptides to large macromolecular systems, integrating data from experiments, statistical analyses, physical principles, and prior models [1, 9, 11, 13]. IMP possesses several advantages over other modeling software. It allows the modeler to combine a wide variety of experimental data. Data from multiple sources can be integrated objectively *via* Bayesian inference. Multi-scale representations allow the modeling of regions of unknown structure alongside regions of known structure. The modular design facilitates mixing and matching of scoring functions and sampling algorithms. The method is scalable for modeling large assemblies such as the nuclear pore complex. Finally, models are thoroughly assessed and validated; this includes generating an ensemble of models consistent with the data, instead of a single best-fit model. However, stochastic sampling in IMP may not produce the native structure and coarse-graining may omit atomistic information such as hydrogen bonding.

Integrative modeling with IMP comprises four stages – data gathering, system representation and translating data into spatial restraints, sampling, and analysis and validation (Figure 1) [1, 11].

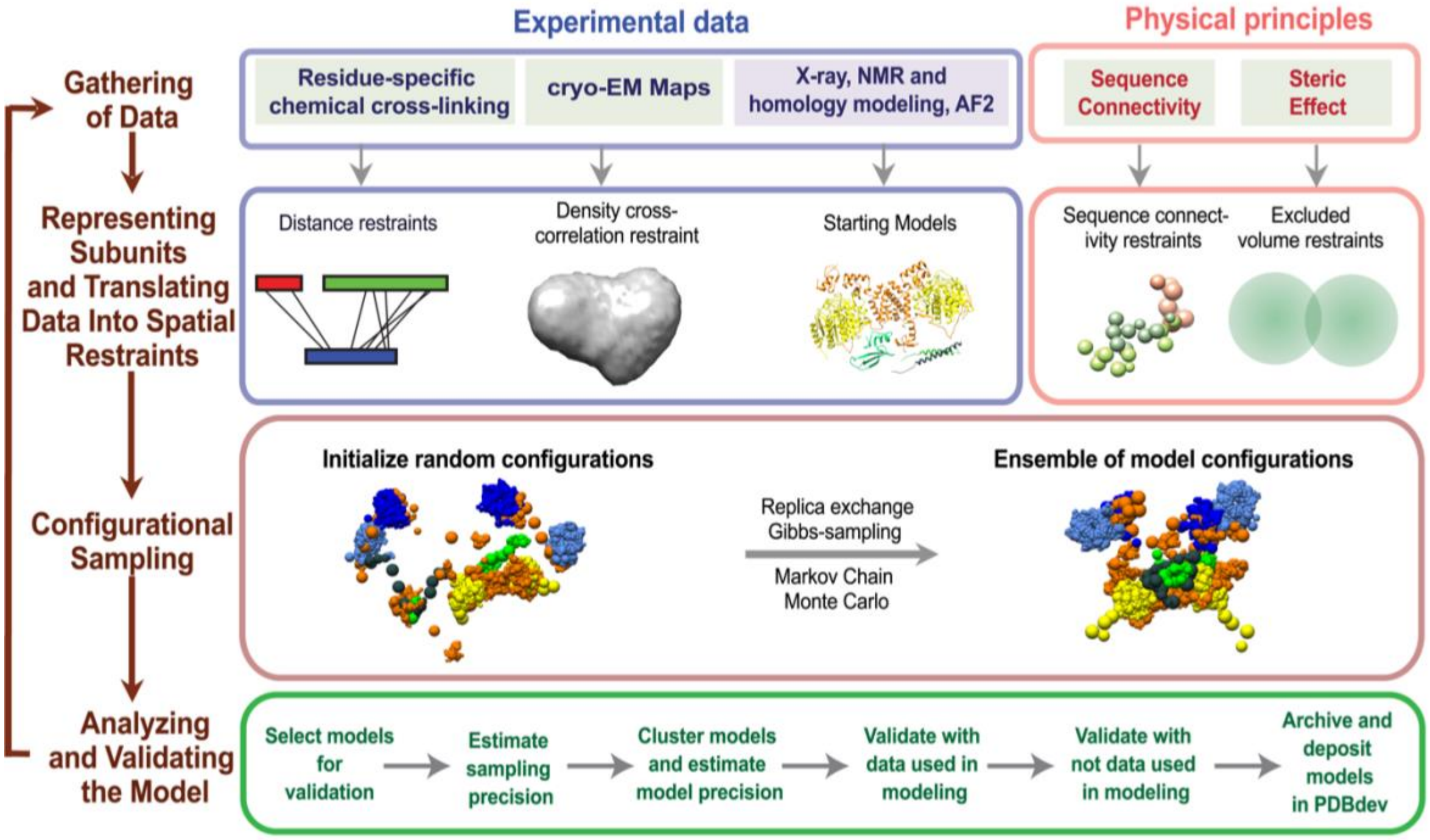

**Fig. 1 Integrative structure determination workflow** Schematic describing the workflow for integrative structure determination. The first row describes the input information. The second row details how data are used to encode spatial restraints. The third row describes the sampling method, and the last row describes the analysis and validation protocol.

## Gathering data

All available information for the system of interest that informs the relative positions and orientations of the subunits of the system can be used for structure determination. Atomic structures of the subunits can be obtained from X-ray crystallography, NMR, homology modeling, and AlphaFold2 (AF2). Data from electron microscopy and small angle scattering (SAXS, SANS) can inform the shape of the complex, whereas data from XLMS, Y2H, co-IP, and FRET can inform residue-level and/or domain-level interactions. A comprehensive list of types of experiments that can be used for integrative modeling is available in [9]. In addition, physical principles like sequence connectivity, excluded volume, and statistical potentials from previous structures are also used. The available data for the system can be used in multiple ways: representing the system, *i.e.,* all-atom or coarse-grained, scoring and sampling models, filtering the models based on fit to input data and model validation [9, 11, 14].

### Representing Subunits and Translating Data into Spatial Restraints

Model representation involves assigning a set of geometrical primitives such as points, spherical beads, cylinders, Gaussians, and probability densities to each subunit in the system [1]. At the highest resolution, this primitive could represent an atom, or in a coarse-grained representation, it could represent a residue or a set of residues. IMP allows for a multi-scale representation where different representations can be employed to describe different parts of the system. The same part can also have multiple representations [1, 10]. Regions of known structure are modeled as a rigid body, where the relative positions of the primitives are constrained, whereas regions for which no structure is available can be modeled as flexible beads [11].

The input data is then translated into spatial restraints that measure deviation from a measurement or theoretical estimate. For example, data on the interaction between two domains can be translated as a harmonic restraint, $d = k(x - x_0)^2$ where, $x_0$ represents the distance between the domains as obtained from experiment and $k$ represents the weight of the restraint. IMP provides restraints for several types of experimental data such as chemical crosslinking, EM, FRET, HDX, and SAXS [15–19]. Stereochemistry restraints such as sequence connectivity and excluded volume are also provided. Apart from this, custom restraints can also be implemented. These restraints form a Bayesian scoring function that quantifies the degree of satisfaction of input information by the model [1, 11].

### Sampling

The initial system configuration is randomized by shuffling the subunits. Usually, models are sampled by Gibbs Sampling with Replica Exchange Markov Chain Monte Carlo algorithm [9, 11]. IMP also provides several other optimization methods and sampling algorithms including conjugate gradients, molecular dynamics, and simulated annealing. Each rigid body is associated with six degrees of freedom - three translational and three rotational, whereas a flexible bead is associated with three translational degrees of freedom. Thus, a system consisting of $R$ rigid bodies and $F$ flexible beads has $6R + 3F$ degrees of freedom. As a result, sampling becomes increasingly expensive for larger systems [1].

### Analysis and Validation

Sampled models are first analyzed to extract two independent sets of good-scoring models. This step involves filtering models based on score equilibration, followed by density-based score clustering using Hierarchical DBSCAN (HDBSCAN), an assumption-free clustering method [11]. An assessment of sampling exhaustiveness is then performed on the two extracted independent model samples. The assessment includes testing for convergence of total score and testing the similarity between the score and structure distributions of the two samples [10]). To assess the similarity of structure distributions, models from both samples are pooled and clustered using a series of distance thresholds. The model precision is computed from this protocol as the average RMSD of the models in a cluster with respect to the cluster centroid model. It provides a lower bound on the model accuracy, which determines the applicability of the model. The output of modeling can be visualized either as a representative cluster centroid model or a localization probability density (LPD) map. The latter specifies the probability of a voxel being occupied by a bead in a set of superposed models; it is similar to an EM map [11]. Output models are validated by the fit of models to input data and data not used in modeling. Input data quality is also validated. Finally, integrative models and data are deposited in the PDB-Dev, a prototype archive of integrative (or hybrid) structures and associated data developed in the Worldwide Protein Data Bank, which will be integrated with the wwPDB (wwPDB) [1, 11, 20, 21].

## Assembline

Assembline is a protocol for the integrative modeling of large macromolecular complexes [8]. It combines Xlink Analyzer, UCSF Chimera, and IMP for modeling. Assembline was developed to make integrative modeling more efficient in cases where a large number of atomic structures need to be fit to EM maps. The workflow involves defining the system representation and scoring function, generating fit libraries, global optimization, recombination, refinement, and analysis. Assembline allows for atomic as well as coarse-grained representations for the system. The input structures are treated as rigid bodies by default, although it provides means to account for conformational flexibility. The input information is translated into spatial restraints which form a scoring function. Assembline uses restraints from IMP along with some custom-defined restraints. However, the generated fit library restricts the search space and can be inaccurate for a low-resolution map. As a result, the sampling in Assembline may be inexhaustive and may not converge to the native structure.

The initial atomic structures are fit to the EM map using the Fit in Map tool of Chimera to generate an ensemble of non-redundant rigid body fits, generating a fit library. Models of the system are generated by sampling fits from the precalculated fit library *via* global optimization using simulated annealing, where, at each step models are scored based on the fit to EM map and other restraints. Next, a recombination step is performed for the enrichment of good-scoring models. This step is optional but useful for systems having several rigid bodies and a large fit library. Herein, the global optimization is performed again using a reduced fit library comprising only fits that led to a good-scoring model in the previous global optimization step. The global optimization steps need not be performed if an initial model for the complex already exists. In such a scenario, one can directly start with the refinement step described next.

The structure(s) generated *via* global optimization and recombination are used for refinement. A coarse-grained representation is chosen. The fit library is no longer used for sampling, rather new conformations are sampled *via* simulated annealing for rigid bodies and conjugate gradients for flexible beads. Models are scored based on restraints from experiments and physical principles. The analysis and validation step follows the IMP workflow described above.

## HADDOCK

High ambiguity-driven docking (HADDOCK) is a protein-protein docking method that uses biochemical and biophysical data obtained from NMR, SAXS, mass spectrometry, and mutagenesis studies used to drive docking [22, 23]. The advantages of HADDOCK include atomic-resolution integrative modeling that incorporates protein flexibility, the ability to use physical force fields along with a limited number of experimental restraints, and an easy-to-use web server interface.

The overall docking procedure can be categorized into three stages. The first stage involves rigid body energy minimization. It begins by separating the partner proteins and randomizing their orientations. This is followed by four rounds of inter-molecular energy minimization where each protein is rotated at random, keeping the other fixed. Subsequently, the energy of the complex is minimized while keeping the two proteins rigid and allowing them to translate and rotate. The second stage involves three semi-rigid simulated annealing steps. In the first step, the proteins are treated as rigid bodies for orientational optimization. In the second step, only the interfaces are allowed to move. Lastly, both the backbone and side chains at the interface are allowed to rotate and translate. The third stage involves molecular dynamics (MD)-based refinement in explicit solvent incorporating restraints from experimental data. Starting with a refinement step in a shell of TIP3P water molecules, the system is heated to 300K and MD is performed at 300 K with restraints on the non-interface heavy atoms. Subsequently in the cooling stages, MD is performed with restraints limited to the non-interface backbone atoms. Pairwise backbone interface RMSD (iRMSD) is used for clustering the final structures.

The experimental data about the interacting residues used to drive the docking procedure is introduced in the form of an ambiguous interaction restraint (AIR). Residues that form intermolecular interactions are used to define AIRs and can be classified as active or passive based on the experimental data. For example, in case of NMR residues showing a significant chemical shift perturbation upon complex formation and having a high solvent accessibility are considered as active while others as passive. AIRs restrict the search space and enable HADDOCK to sample all possible configurations around the interacting site.

Thus, HADDOCK allows for efficient sampling of configurations at the interface and enables modeling at high resolution driven by a physical force field with experimental data incorporated as restraints. However, due to its use of an atomic representation, it cannot be feasibly applied to large assemblies like NPC [8]. Also, it requires prior knowledge of the binding site and does not efficiently sample EM maps [8].

## Integrative Structure Determination using Density Maps

Integrative modeling also encompasses methods that involve *de novo* model building (*e.g.*, Phenix, Coot, DeepTracer) and fitting of structures into low-resolution density maps (*e.g.*, TEMPy-REFF, MDFF, DeepMainMAST). Coot is a molecular graphics tool that allows for visualization and model manipulation including idealization, real-space refinement, manual rotation and translation, rigid-body fitting, ligand search, solvation, mutations, and Ramachandran idealization [24, 25]. Similarly Phenix was developed for automating macromolecular structure determination [26]. It also provides tools for experimental phasing, molecular replacement, automated model building, refinement, validation, and deposition. Recently, deep learning-based methods have been employed for *de novo* model building. DeepTracer uses a deep convolutional neural network along with the cryo-EM map and amino acid sequence to output the atomic structure for protein complexes [27]. [28] developed DeepTracer-ID which improves upon DeepTracer by incorporating AlphaFold-multimer predictions. In this approach, DeepTracer is used to first generate the backbone trace from a cryo-EM map. This is followed by aligning the backbone trace with the AlphaFold predicted models.

Next, we discuss methods for fitting structures into density maps. One approach is to use MD simulations to fit existing structures to density maps. [29] develop a method for flexible fitting of atomic structure into low-resolution cryo-EM maps using MD simulations driven by an additional effective potential which is proportional to the cross-correlation between the cryo-EM map and the density map generated from the atomic structure. It aims to maximize the cross-correlation in order to obtain a better fit of the atomic structures to the cryo-EM density map. In Molecular dynamics flexible fitting (MDFF), on the other hand, the additional potential is derived directly from the cryo-EM map [30, 31].

A recent method, TEMPy-REFF (REsponsibility based Flexible-Fitting), uses mixture models for atomic structure refinement in cryo-EM density maps [32]. Using an Expectation Maximization algorithm, it estimates the atomic positions and local B-factors. Given an initial set of atomic coordinates, their associated B-factors, and resolution, a simulated density map and map-derived forces are computed and co-ordinates and B-factors are updated based on the forces. It also computes the ensemble representation of the models given the atomic B-factors.

Terashi G. et. al. (2023) develop a deep learning method for protein structure modeling using cryo-EM density maps, augmented with AF2 predictions. Their approach, DeepMainmast, is a deep learning method that traces protein chains de novo given the cryoEM map and the sequences of the proteins involved. First, the probabilities of twenty amino acids and six atom types (N, C, C$\alpha$, C$\beta$, O, and others) are computed at every grid point in the density map using Emap2sf which involves a UNet architecture. The grid points with high C$\alpha$ probability are then clustered using mean shift clustering to generate Local Dense Points (LDPs). Second, the LDPs are connected to generate C$\alpha$ paths using a Vehicle Routing Problem (VRP) solver algorithm. Third, the computed C$\alpha$ paths are aligned with the target sequence using the Smith-Waterman dynamic programming algorithm. The matching score for the Cα positions and the amino acid type in the alignment is defined by the DAQ(AA) score [33] obtained from Emap2sf output. This typically generates thousands of Cα path-sequence alignments per protein chain, which are referred to as Cα fragments. Regions of the AF2 model, with an RMSD less than 1.5 Å for nine residues or longer, with the Cα fragments by DeepMainmast are added to the set of fragments. 54 such fragment libraries are constructed using different parameter combinations. Fragments from each of the Cα fragment libraries are then assembled using a Constraint Programming (CP) solver to maximize the DAQ score [33] to build 54 Cα models. The generated Cα models are further fragmented into ten-residue long Cα fragments which are reassembled into three new Cα models by the CP solver. In addition to being used for supplementing the fragment library, the AF2 model is also superposed on the density map using VESPER [34]. Simulated maps from the AF2 model are generated at three different resolutions, with ten superpositions computed for each. One final Cα model is constructed by assembling the thirty superpositions. This produces a total of four Cα models. For all-atom model building, the Cα model is subjected to PULCHRA [35] followed by using RosettaCM [36] for filling the missing regions and refinement. A total of 20 models are generated from four Cα models. DAQ score and DOT score are used for evaluating and ranking these models with the final score being a sum of the two scores [32–34].

[37]develop cryoFold for generating ensembles of protein structures at atomic resolution using protein sequence and cryo-EM data. It combines three methods including MAINMAST [38], resolution-exchange molecular dynamics flexible fitting (ReMDFF) [39], and modeling employing limited data (MELD) [40]. Given a cryo-EM map with 5 Å or lower resolution, MAINMAST is used to generate a backbone tracing. This serves as an input for iterative refinement using MELD, which samples plausible secondary structures, and ReMDFF, which fits the protein backbone and side chain conformations into the cryo-EM density map for each MELD-generated search model.

## Dynamical structure modeling

In contrast to static structure modeling, dynamical modeling captures the small and large-scale fluctuations of a system, for example, multiple states, dynamic interactions, and flexibility of loop regions. Dynamical modeling is essential for understanding biological phenomena, such as protein folding, molecular interactions, and cellular processes, where temporal aspects play a crucial role in shaping the system's functionality. Here we describe integrative modeling methods for dynamical data.

### Plumed Integrative Structural and Dynamical Biology (PLUMED-ISDB)

PLUMED-ISDB is a package for integrative dynamical modeling of heterogeneous systems [41]. It implements the metainference framework which allows for integrating noisy, ensemble-averaged experimental data from NMR, cryo-EM, FRET, and SAXS [42]. It is implemented as a module for the PLUMED package, a suite of tools for free energy calculation using metadynamics [41] [43, 44]. PLUMED-ISDB facilitates the integration of a wide range of experimental data, enabling its utilization in molecular dynamics simulations for both sampling and refinement.

### BioEn

Bayesian Inference of Ensembles (BioEn) is a method for ensemble refinement based on Ensemble Refinement for SAXS (EROS [45]) and Bayesian replica ensemble refinement [46]. EROS is a method that uses coarse-grained models to generate an initial pool of conformations which are then refined for fit to SAXS data [45]. However, it does not scale well with the number of structures. Moreover, the resulting structures may not be representative of the refined ensemble if the initial pool of configurations does not overlap with the optimized ensemble [46]. Bayesian replica ensemble refinement uses replica exchange with Bayesian ensemble-averaged restraints for ensemble refinement. However, it becomes computationally expensive with a large number of replicas. BioEn combines the two methods, using a small number of replicas to obtain an enriched pool of configurations sampled with Bayesian replica ensemble refinement, for which the optimal set of weights are determined using EROS. This approach is more efficient, allowing faster convergence towards the optimal Bayesian ensemble, as the EROS reweighting is less expensive compared to running simulations with a large number of replicas.

# Recent Applications of Integrative Structure Modeling

In this section, we discuss some of the recent applications of integrative structural modeling to characterize macromolecular assemblies involved in chromatin remodeling, transport across membranes, cell motility, cell-cell adhesion, and immune response.

## Chromatin-modifying complexes

Macromolecular assemblies that are involved in modulating chromatin architecture play a role in several processes such as modulating transcription, translation and DNA damage response, maintaining genome integrity, thereby regulating cell division and differentiation [47]. Some recent examples of these assemblies characterized using integrative structural modeling are described here.

## WDR76-SPIN1-nucleosome complex

The WD40 repeat-containing protein, WDR76, is a histone-binding protein. It is involved in DNA damage response, transcriptional regulation and ubiquitination [48–62]. Moreover, mutations in WDR76 are linked to colon and lung cancers in humans [63–65].

[57] utilized integrative structure modeling to characterize the WDR76-SPIN1-nucleosome complex using the Integrative Modeling Platform (IMP, https://www.integrativemodeling.org, [9, 13, 66]) relying on structures from X-ray diffraction [67] and AlphaFold2 [68], data from SCAP-XL (Serial Capture Affinity Purification along with XLMS), and stereochemistry considerations. In the integrative structure, they showed that WDR76 and SPIN1 simultaneously interact with the same H3 histone tail. Based on the structure, they proposed a model for the assembly and function of the WDR76-SPIN1 complex in which SPIN1 first binds to the target nucleosome, and then recruits WDR76. The WD40 domains of WDR76 may subsequently scaffold larger protein complexes at the target nucleosome [57].

## Rvb1-Rvb2-INO80 complex

Rvb1 and Rvb2 (Rvbs) from *S. cerevisiae* are two members of the highly conserved AAA+ (ATPases Associated with diverse cellular Activities) family of proteins. They play a crucial role in transcription, small nucleolar ribonucleo-protein assembly, DNA-damage repair, signaling, cell division, and differentiation. It was previously known that these Rvbs form hexamers and dodecamers [69]. Based on their structural similarity with the bacterial helicase RuvB, Rvbs were thought to have a DNA helicase activity. In addition, it was also proposed that they chaperone protein assemblies [70]. However, their mechanism of action is poorly understood [70, 71].

To investigate the role of Rvbs as protein assembly chaperones, [71] used integrative structure determination using IMP (https://www.integrativemodeling.org, [9, 13, 66]) relying on data from SEC-MALS, native MS, XLMS, cryo-EM, X-ray crystallography, comparative modeling, secondary structure predictions, and stereochemistry considerations [71–73]. Based on the data from SEC-MALS and native MS experiments, a 6:6:2 (Rvb1:Rvb2:Ino80) stoichiometry was used for modeling the dodecameric assembly. Integrative modeling resulted in three clusters, each of which matched the observed cryo-EM density and satisfied most of the input crosslinks. Since all three clusters satisfied the input information and had similar populations, the authors inferred that the dodecamer is conformationally heterogeneous. They also showed that the two Ino80s bind asymmetrically to the barrel-shaped dodecamer formed by the Rvbs [71].

Based on the integrative structure, the authors proposed a model for the protein assembly chaperone activity of Rvbs. In their model, Rvbs, upon binding to a nucleotide, fail to dodecamerize. Binding to a client, such as Ino80, facilitates dodecamerization of Rvbs. Upon addition of ATP, the dodecamer splits into two hexamers releasing the client to be assembled in a large assembly [71].

## Smc5/6 complex

Structural Maintenance of Chromosomes (SMC) complexes play a crucial role for regulating genome organization in both prokaryotic and eukaryotic cells. In general, the SMC complexes have two core SMC subunits and a set of non-SMC subunits. The core SMC subunits usually have an elongated architecture with the N-terminal and C-terminal ATPase domains close together, forming the “head” region, and an elongated “arm” region which folds on itself at the “hinge” region and has sharp bends called the “elbow” sites [74, 75]. Two such core SMC subunits form a lateral dimer which serve a structural role and form the backbone of the complex. The non-SMC subunits impart specific functions, such as providing target site specificity. The Smc5/6 complex regulates DNA replication and recombinational DNA damage repair [76–78]. In addition to Smc5 and Smc6, it also comprises Nse1 through Nse6 [79–82]. Relying on data from X-ray crystallography, cryo-EM, negative staining electron microscopy (NS-EM), XLMS, and structures obtained from comparative modeling (for head and hinge regions of Smc5 and Smc6), parametrically designed coiled-coil (for the Smc5/6 arm regions), and stereochemistry considerations, [83] determined the molecular architecture of a five subunit sub-complex comprising Smc5, Smc6, Nse2, Nse5, and Nse6 using IMP (https://www.integrativemodeling.org, [9, 13, 66]). The integrative structure shows that, in contrast to the canonical SMC complexes such as cohesin and condensin, in Smc5/6, the “arm” regions do not bend at the “elbow” sites, keeping the “hinge” away from the “head” region. Overall, their integrative structure elucidated the unique structural and functional properties of the Smc5/6 complex.

## NuRD complex

Nucleosome Remodeling and Deacetylase (NuRD) complex is a chromatin-modifying assembly that comprises a histone deacetylase module and a chromatin remodeling module [47, 84, 85].

The deacetylase module comprises scaffold proteins (MTA1/2/3), histone deacetylases (HDAC1/2), histone chaperones (RBBP4/7), methyl-CpG binding proteins (MBD2/3) and GATA-type zinc-finger proteins (GATAD2A/B). Mutations in these protein subunits are implicated in various cancers and developmental disorders [47, 86]. Previously, limited structural information was available for the NuRD complex. For three of its subcomplexes (MHR - MTA1:HDAC1:RBBP4, MHM - MTA1$^{N}$:HDAC1:MBD3:GATAD2A$^{CC}$, and NuDe - MTA1:HDAC1:RBBP4:MBD3:GATAD2A$^{CC}$), the stoichiometry, low-resolution EM maps, atomic structures of a few domains of the subunits, and XLMS data were known [87–93].

[94] determined the integrative structures of these three subcomplexes of the NuRD complex relying on information from experiments such as SEC-MALS, DIA-MS, XLMS, NS-EM, X-ray crystallography, NMR spectroscopy, homology modeling, secondary structure predictions, and stereochemistry considerations using IMP (https://www.integrativemodeling.org), [9, 13, 66]). The integrative structures for the three sub-complexes were in agreement with the input XLMS and EM data. They were also corroborated by data from independent experiments such as XLMS and cryo-EM [95–98].

They identified novel protein-protein interfaces in the NuDe sub-complex which can be experimentally verified. Further, based on the integrative structures of the three subcomplexes, the authors proposed a two-state model for the binding of MBD3 in NuRD. They proposed that in the absence of MBD3, the C-terminal arms of MTA1 in MHR are flexible, facilitating interaction with nucleosomes and transcription factors. In this configuration, MBD3 can interact with the MTA1 dimerization interface. It then recruits the chromatin remodeling module and shifts away from the MTA1 dimerization interface. This model proposes that the primary role of MBD3 is in connecting the two NuRD modules rather than binding to DNA, which is supported by other studies [99, 100]. Overall, they provided the most complete and accurate structure of the NuRD complex so far [94].

## Ciliary and flagellar complexes

Mammalian sperm flagella consist of nine doublet microtubules (DMT), and endpiece singlet microtubules (SMT) which are stabilized by the Microtubule Inner Proteins (MIPs). Using cryo-ET and AlphaFold2 predictions, [101, 102] identified the MIPs in mammalian sperm flagella. [101] identified Tektin 5, CCDC105 and SPACA9 as novel MIPs which were validated by biochemical assays and mass spectrometry. [102] identified MIPs between the DMTs and SMTs and determined their molecular organization.

Similar studies involving cryo-EM structure determination combining AlphaFold2 and AlphaFold-multimer predictions, followed by model refinement using Phenix and Coot have been conducted to determine the organization of the intraflagellar transport trains [103], and the ciliary central apparatus [104].

# Cell-cell junctions

Cell-cell junctions, such as desmosomes, are assemblies that connect the cytoskeleton of adjacent cells. Desmosomes are abundant in tissues under mechanical stress such as heart and epithelial tissue [105]. Recently, an integrative modeling study combined data from X-ray crystallography, electron cryo-tomography, immuno-electron microscopy, yeast two-hybrid experiments,co-immuno precipitation, *in vitro* overlay, *in vivo* co-localization assays, in-silico sequence-based predictions of transmembrane and disordered regions, homology modeling, and stereochemistry information to produce an integrative structure of the outer dense plaque of desmosomes [106]. The structure refined known protein-protein binding sites, rationalized disease mutations, and provided insights on the localization and function of disordered domains of Plakophilin and Plakoglobin. Another study on the extra-cellular region of desmosomes combined data from electron cryo-tomograms with flexible fitting of known structures using MD simulations to understand the mechanistic basis for the plasticity of cadherins [107]. This explains the robustness of desmosomes to mechanical stress.

# Membrane-trafficking complexes

## Nuclear pore complexes

The nuclear pore complex (NPC) is one of the largest assemblies in the human cell. It is embedded in the nuclear membrane and regulates the flow of DNA, RNA, and protein between the across the nuclear membrane. The past three years witnessed great advances in the structure of the NPC. We first discuss three studies that characterized the cytoplasmic and linker-scaffold regions of the NPC based on rigid fitting of Alphafold and X-ray structures to medium-resolution maps and cryo-electron tomograms. Next, we detail two studies that resulted in a comprehensive integrative structural and functional characterization of the yeast NPC. Finally, we discuss multi-state integrative models of the NPC and the integrative model of the assembly pathway of the NPC.

First, the cytoplasmic ring (CR) of the *Xenopus laevis* NPC was elucidated by single-particle cryo-EM on intact NPC from *Xenopus* oocytes, reaching a medium resolution of 5-8 Å [108, 109]. Near-atomic resolution integrative structures of the CR were determined based on rigid fitting of Alphafold models in cryo-EM maps and validated by biochemical studies and chemical crosslinks. They revealed novel interfaces between the nucleoporins of the CR and identified the protein in the bridge domain of the CR.

Second, the integrative structure of the cytoplasmic face of the human NPC was based on rigid fitting of a new X-ray structure of Nup358, the cytoplasmic filament nucleoporin (CFNC), to previous cryo-electron tomograms of the NPC. The structure was validated by cell-based assays and biochemical studies. It revealed the symmetry of the cytoplasmic face, which is surrounded by Nup358-Ran complexes on the outside and coat nucleoporin complexes on the inner face of the central transport channel [110]. The outer ring Nup93 was shown to anchor the CFNC to the cytoplasmic face. The function of Nup358, the CFNC, was determined to be linked to translation.

Third, the integrative structures of the linker-scaffold network of the inner ring of the human and yeast NPCs were determined by X-ray and cryo-EM structures of Nup188 and Nup192 bound to Nic96, Nup145N, and Nup53, cryo-ET maps, and biochemical studies. The structures revealed that the linker-scaffold interactions induce plasticity and robustness in the inner ring, whereas they have an opposite effect in the outer ring, making it more rigid. The linker scaffold is conserved despite significant sequence divergence [111].

A composite integrative multi-scale structural and functional characterization of the yeast NPC was performed in two studies [112, 113]. A new cryo-EM map of the inner ring of the NPC and a new cryo-electron tomogram of the NPC were integrated with previous EM maps, chemical crosslinks, quantitative fluorescence imaging, and biochemical studies to obtain the integrative multi-scale structure of the yeast NPC. The studies revealed the architecture of the core scaffold with flexible connectors between layers. A modular arrangement of protein complexes was observed. The architecture of connectors in the central channel outlined mechanisms by which transport *via* the NPC could be modulated. Mechanistic insights into the plasticity of the pore, including the role of Pom153 in ring dilation, were uncovered.

Multiple states of the NPC were also modeled. Integrative structures of the constricted and dilated states of the human NPC were obtained based on cryo-electron tomograms of the NPC in these states and structural models from Alphafold2. The structures were validated by biochemical studies [114]. They revealed the binding sites for disordered nucleoporins and the conformational changes that occur during dilation and constriction of the pore.

Finally, integrative spatiotemporal models of the assembly pathway of the NPC were obtained based on fluorescence correlation spectroscopy (FCS)-based live imaging of nucleoporins [115]. The model revealed the molecular mechanisms underlying two separate pathways of assembly and characterized the structures of assembly intermediates. Taken together, all these advances were a result of both improved imaging techniques and recent deep learning-based methods for structure prediction.

### Type III Secretion System

The type III secretion system (T3SS) is a mega-dalton cell-surface macromolecular machine that facilitates the entry of virulence factors into the host cell during infection by pathogenic bacteria. It is an important drug target due to its role in infection. Detailed structural characterization, by cryo-EM, has been performed for the *Salmonella* T3SS, but little is known about the T3SS structures in other species. A partial cryo-EM structure of the *Shigella* T3SS was available but lacked several details such as the structure of the export apparatus, and the interaction between the export machinery and the inner membrane ring. To fill in these gaps and obtain a more complete structure, integrative structure determination of the *Shigella* T3SS was performed based on *ab-initio* reconstructions of subunit from the *Shigella* T3SS cryo-electron map and chemical crosslinks [116]. The structure allowed for the characterization of similarities (conserved features) and differences (unique features) of the T3SS from different species. Like other T3SSs, the *Shigella* T3SS architecture consists of the export apparatus anchored in the membrane-embedded needle complex. Uniquely, the Shigella T3SS contains a new fold of the C-terminal S domain of the secretin pore complex and a new subunit, SpaS, in the export apparatus.

## Complexes involved in immune response

Nanobodies (Nbs) have become a cost-effective alternative for use as therapeutics due to their higher solubility and stability compared to antibodies [117, 118]. A nanobody is the $V_HH$ domain from homo-dimeric heavy chain-only functional antibodies (hcAbs) found in camelid sera [119].

Two structure-prediction methods were developed recently for nanobodies: one to the predict structures of nanobodies from their sequence and another to characterize the structures of nanobody-antigen complexes.

NanoNet predicts the structure (coordinates of backbone and Cβ atoms) of given the sequence of a Nb, the VH domain of a monoclonal antibody, or the Vβ domain of a T-cell receptor (TCR) using convolutional neural networks (CNNs) comprising of 1D residual network (ResNet) blocks to [117]. (Xiang et al 2021) used comparative modeling and loop refinement followed by structural docking with the antigen to obtain structures of nanobody-antigen complexes. XLMS and mutagenesis studies were used to validate structural models for these complexes. This was used as part of a pipeline for global identification, classification, and high-throughput structural characterization of Nbs, given an antigen. It enabled the identification of 8000 high-affinity Nbs for the receptor-binding domain (RBD) of the SARS CoV2 spike glycoprotein [118, 120].

[121] developed a high-throughput computational pipeline based on structure prediction and docking to obtain structures of antibody-antigen complexes. Structures of antibody and antigen were generated by structure predictors like ABodyBuilder2 [122] and AF2 [123], respectively. HADDOCK3 was used for docking with the AIRs defined using experimentally-determined interface residues on the antibody and antigen. In contrast to the default HADDOCK protocol, where thousands of models are generated in the initial rigid body docking phase, it was shown that tens of models are sufficient. This protocol for modeling antibody-antigen complexes was shown to outperform AF2-multimer.

# Recent Advances in Integrative Modeling Methods

Here, we discuss recent advances in integrative modeling including new sources of data, new computational methods for integrating experimental data such as EM and chemical crosslinks, advances in model representation, sampling, and analysis.

## New sources of data

A genetic interaction quantifies the relationship between two or more genes that affect the phenotype of an organism [124]. *In vivo* genetic interaction measurements were incorporated as a new source of data for integrative modeling [125]. Several point mutations were made on each protein in a complex. Phenotypic profiles were constructed for each mutation by crossing with thousands of gene deletions and environmental perturbations. The similarity between phenotypic profiles of mutations in two proteins indicates their spatial proximity in the complex. This similarity was quantified in a score that was translated to a Bayesian upper bound distance restraint between mutated residues. Genetic interaction data was used in this manner to determine the integrative structure of the H3-H4 complex [126]. This type of interaction data provides the same accuracy and precision as chemical crosslinks and can be especially useful for localizing disordered regions. It is complementary to structure predictions from AI-based methods such as AF2 [126].

Another emerging source of data is from native cell extracts [127]. The advantage of native cell extracts is that they can provide *in situ* structures of protein communities, defined as higher-order, transiently interacting, sets of macromolecular complexes that come together spatially to perform cellular functions. For example, metabolons and enzymatic pathways form communities. In contrast to highly purified *in vitro* constructs, native cell extracts can enable *in situ* identification of unstable and transient complexes, abundance and dynamics of individual complexes, and interactions between complexes. Experiments on cell extracts include proteomics to determine the identities of proteins in these complexes, chemical crosslinking to identify residue-level interactions, and cryo-EM on extracted complexes to characterize *in situ* structures. Since the proteins can be easily identified within these extracts, integrative structure determination based on native cell extracts is complementary to cryo-electron tomography which provides *in situ* structures but may not be able to determine protein identities (See Frontier 2). Recent integrative structural studies on cell extracts from thermophilic bacteria have shed light on the structure, dynamics, and interactions of several enzymatic complexes within distinct protein communities. Multiple states and binders were identified for a fatty acyl synthase from thermophilic bacteria [127]. Studies on the oxoglutarate dehydrogenase complex produced a high-resolution cryo-EM structure of its core protein (E2o), determined the interactions of E2o with two other enzymes in the same pathway, and identified a new subunit connecting two enzymes in the pathway [128]. A cryo-EM structure of the active state of a pyruvate dehydrogenase complex core and integrative structures of the entire 10 Mda pyruvate dehydrogenase complex from thermophilic bacteria based on cryo-EM, quantitative mass spectrometry, crosslinking mass spectrometry, and computational models provided mechanistic insights into the transacetylase reaction and organization of the reaction in distinct cellular compartments, and helped propose minimum reaction paths among enzymes [129, 130].

Recent AI-based protein structure prediction methods, such as AF2, are trained on a database of high-resolution X-ray and cryo-EM structures [68]. AF2 predicts a single structure, and does not predict structural heterogeneity. Therefore, its predictions on intrinsically disordered regions (IDRs), which lack a single well-defined structure and exist in an ensemble of conformations, are usually not reliable. They are involved in important cellular functions such as molecular recognition, signaling, gene regulation due to their ability to bind to multiple partners, overcome steric restrictions, and undergo induced folding [131]. They are also present in large macromolecular assemblies, for example the Fg repeats in Nucleoporins which are a part of the Nuclear Pore Complex (NPC) [113]. Several recent integrative modeling studies have highlighted the importance of IDRs in the assembly, for example, the MBD3 IDR in the NuRD complex [94] and the Plakophilin-head region in desmosomes [106]. More data on disordered regions focused on these will improve their structural characterization in macromolecular assemblies.

# Improved methods for modeling with EM and XLMS data

Due to continuing advances in instrumentation and analysis algorithms, cryo-electron microscopy (cryo-EM) and chemical crosslinks from mass spectrometry (XLMS) have revolutionized the structural characterization of large macromolecular complexes. They are expected to be a major source of data for integrative modeling as they continue to evolve. Here, we discuss recent methods for structural characterization using data from cryo-EM and XLMS.

## New methods for Electron-Microscopy (EM) data

Owing to the recent advances in instrumentation and in the image processing algorithms, cryo-EM has proven to be an invaluable tool for structural characterization of macromolecules [132, 133]. However, building accurate structural models relying on cryo-EM data is challenging due to the noise in the data, radiation damage to the sample upon prolonged exposure and the limitations of the current image classification, 3D reconstruction and model building techniques [15]. First, we discuss a deep learning method for *ab initio* characterization of protein secondary structure and nucleic acids in cryo-EM maps. Next, we discuss two methods for characterizing heterogeneity in structures from cryo-EM maps. Next, we discuss an advancement to Bayesian modeling of EM data. Finally, we discuss improvements to validating models from cryo-EM maps.

**Emap2sec+** [134] recently developed a deep learning-based method, Emap2sec+, to detect nucleic acids (DNA and RNA) and secondary structures of proteins in cryo-EM maps of 5-10 Å resolution. Emap2sec+ predicts the probability for a secondary structure in a 11 $Å^3$ voxel subvolume of the input cryo-EM map. It does so in two phases: In the first phase, it passes the 11 $Å^3$ query voxel through five CNN (Convolutional Neural Network) models, four of which perform binary classification to predict the probability of a protein secondary structure ($\alpha$ helix, $\beta$ sheet or others) or a nucleic acid being present in the voxel. The fifth model performs a multi-class classification for the above classes ($\alpha$ helix, $\beta$ sheet, other protein secondary structure or nucleic acid). The second phase takes the eight probabilities for the query voxel and $7^3$ neighboring voxels as input and passes them through another deep CNN and predicts probabilities for the above four classes. Emap2sec+ significantly improves the accuracy of secondary structure and nucleotide prediction over previous methods. Nucleotides, in particular, are predicted accurately even for low-resolution maps.

**DynaMight** Several existing cryo-EM workflows fail to capture structural heterogeneity in macromolecular regions that undergo continuous structural changes [135, 136]. To structurally characterize such regions, methods such as multi-body refinement and variational auto-encoders (VAE) have been used. Multi-body refinement divides the macromolecular complex into independently moving rigid bodies and performs independent image alignment and reconstruction for each of the individual bodies. VAEs map particle images into a continuous multidimensional latent space that can then be used to reconstruct structures for each point in the latent space [136–142]. Multi-body refinement generates maps that are better resolved for the dynamic regions than the generalized reconstruction approach which does not take structural variability into account. However, it is limited to macromolecular assemblies that have large rigid components. The previous VAE-based approaches, while useful for exploring molecular motions, do not lead to considerable improvements in reconstructed densities for the dynamic regions [136].

[136] improve upon previous VAE-based methods to estimate the conformational variability in cryo-EM datasets. Their approach takes as input two independent sets of particle images (half-set). First, a three-dimensional half map is reconstructed from each set of particle images. Each half map is then used to generate a Gaussian pseudo-atomic model. Each half-set is processed by a separate VAE. The encoder takes individual particle images from the corresponding half-set as input and returns the latent space encoding. The decoder network, using these latent space encodings and corresponding pseudo-atomic models as input, generates a three-dimensional deformation field for the pseudo-atomic Gaussian particles. The VAE is trained using a loss

function that compares the projections of deformed pseudo-atomic models with the experimental particle images. Finally, the deformation fields are then used to reconstruct an improved three-dimensional map. The advantages of this approach include an improved reconstruction procedure using the deformation fields and the estimation of uncertainty in the reconstructions using two independent VAEs. The challenges that remain to be explored include better ways to regularize the model to prevent overfitting.

**cryoDRGN** [141, 143, 144] developed cryoDRGN (cryo Deep Reconstructing Generative Networks) to reconstruct continuous distributions of three-dimensional density maps from cryo-EM datasets of structurally heterogeneous macromolecules. Their method uses an image-encoder-volume-decoder architecture. The encoder produces latent space encodings for two-dimensional particle images. The decoder generates the corresponding Fourier space representation of the three-dimensional density maps from the latent space encodings and positionally encoded three-dimensional coordinates, maximizing the likelihood over particle orientations using a branch-and-bound approach. The advantages of the method are that it can model discrete conformational states as well as continuous trajectories. The latent space encoding can also be used to filter out undesired states. They demonstrate that distinct regions in latent space with high particle densities may correspond to different discrete conformational or compositional states, whereas the latent space regions with low particle densities may correspond to continuous structural transitions. However, the dependence of the pose estimates in cryoDRGN on the traditional consensus 3D reconstructions limits its applicability on datasets with high structural heterogeneity.

**EM Restraint** [15] developed a Bayesian approach for structural characterization of macromolecular assemblies relying on data from low-resolution EM maps, and quantification of noise in the data. Their workflow follows an integrative modeling approach and is implemented in the open-source Integrative Modeling Platform (IMP, https://integrativemodeling.org) [66, 145]. First, a Gaussian Mixture Model (GMM) representation of the input EM map (data-GMM) is generated using a divide-and-conquer algorithm. The structural model for the macromolecular assembly is represented by coarse-grained spherical beads consisting of one or more contiguous residues. Model GMMs based on the beads are generated to describe the EM density of the model and are used to compute the fit of the model to the input EM map. The Bayesian scoring function for the spatial restraint based on the EM data is:

$$S(X) = k_B T. \left\{ - \log p(X) + \frac{N_D}{2} \log \left[ \sum_{k=1}^{N_D} \log^2 \left( \frac{OV_{MD,k}}{OV_{DD,k}} \right) \right] \right\}$$

Where, $S(X)$ is the score (measure of the data violation by the model), $p(X)$ is the prior associated with the model, $N_D$ is the number of components of the data-GMM, $OV_{MD,k}$ corresponds to the overlap between the model and the $k$-th component of the data-GMM, and $OV_{DD,k}$ corresponds to the overlap between the $k$-th component and the entire data-GMM. This approach provides a computationally efficient and smooth approximation for EM data and can handle atomic and coarse-grained models. The correlation in information content of neighboring voxels is accounted for. However, positional ambiguity, highly crowded environments pose challenges for the method. Another limitation for the method is that all objects have an observable density.

**EMMIVox** [132] proposed a Bayesian inference approach, EMMIVox for refining single structure-models, Martini coarse-grained single-structure models, as well as structural ensembles of large macromolecular assemblies from cryo-EM maps. The Bayesian inference framework of EMMIVox generates structural models and accurate B-factor estimates by balancing the experimental data with prior physico-chemical models of the system and its surroundings while explicitly accounting for the correlation and noise in the data. Specifically, correlation between neighboring data points in the map is reduced by pre-filtering voxels, prior models of noise in the map are obtained from two independent 3D reconstructions (half-maps), and per-voxel B-factors that weigh the restraints from each voxel are estimated by a Bayesian inference approach. The conformational ensembles obtained from EMMIVox can then be used for describing the heterogeneity in the low resolution regions of the cryo-EM maps. It is implemented in the PLUMED open-source package [43, 44].

## New methods for chemical crosslinks

Chemical crosslinking by mass spectrometry (XLMS) is an important source of information for integrative modeling and can be performed both on *in vitro* and *in situ* constructs. It relies on the use of a chemical crosslinker which consists of two reactive groups that can bind covalently to the amino acid residues in proteins [146] . These reactive groups are separated by a spacer which defines the length of the crosslinker. Commonly used crosslinkers include disuccinimidyl sulfoxide (DSSO) and 4-(4,6-dimethoxy-1,3,5-triazin-2-yl)-4-methylmorpholinium (DMTMM) [146, 147]. The data from XLMS consists of a set of upper-bound distances between crosslinked residues in the complex. This provides medium-resolution (20-40 Å) information on the structure of the complex and is especially useful when the structures of monomers in the complex are not known. It has been used in several recent integrative modeling studies [94, 113, 114].

Integrative modeling was made more accessible to mass spectrometry experts by integrating IMP with MS Studio, allowing for integrative modeling with crosslinks, EM maps, and known structures from within the mass spectrometry software [148]. Additionally, crosslinks were weighted by the flexibility of the corresponding regions. Data from hydrogen deuterium exchange-mass spectrometry (HDXMS) was used to inform the flexibility of protein regions and classify them into "flexibility zones". Separate Bayesian crosslink restraints were formulated for each zone; more flexible regions were restrained by longer average linker lengths and larger crosslink uncertainties. This approach allows flexible regions to be modeled without introducing unnecessary structuring.

MNXL, a commonly used scoring function for assessing models based on their fit to chemical crosslinks, was extended to assess models of protein complexes [149]. It is a linear combination of the number of crosslinks that violate the maximum upper distance bound of the crosslinker, the expected solvent-accessible crosslinked distance, and the number of non-accessible inter-protein and intra-protein crosslinked residues. The crosslinks score was also combined with a score based on fit to EM maps, to facilitate assessment of models based on multiple sources of data. In a subsequent study, the fit of a model to monolinks was added as a term to the crosslinks score [150]. Monolinks are obtained from the same XLMS experiment and provide information on the surface exposure of a residue. Combining crosslinks and monolinks was shown to result in improved accuracy of scoring.

In an application of deep learning on data from XLMS, a given set of crosslinked residues and input structures was used to predict a narrower distance range for each crosslink using a geometric deep learning approach (XlinkNet) [151]. Although the accuracy of prediction for inter-protein crosslinks is lower than that of intra-protein crosslinks, the narrower distance range (*e.g.*, 12-16 Å) can be used to define more precise spatial restraints for integrative protein-protein docking.

## Advances in model representation

Integrative models can be simultaneously encoded in a variety of representations including one or more of these alternatives: atomic, multi-scale coarse-grained, multi-state, time-ordered (trajectory), and ensemble representations [20, 152]. Atomic representations consist of the Euclidean coordinates of atoms in the system, whereas, in a coarse-grained representation, geometric primitives such as spherical beads or Gaussians are used to represent a group of atoms. A commonly used coarse-grained representation is one where each bead represents a set of contiguous (in sequence) residues of a subunit. Different parts of the system can simultaneously be represented at different scales of coarse-graining. Further, the same region can be coarse-grained at multiple scales simultaneously. This multi-scale coarse-graining approach facilitates efficient sampling and accurate mapping of input information at different resolutions to the model. For instance, chemical crosslinks can be mapped to residue-level beads, whereas low-resolution EM maps can be mapped to ten-residue beads. Often, a set of discrete states is required to represent a system. In this case, a multi-state representation is used, where the model is represented by a set of states and their weights. Further, integrative modeling of the dynamics of a system requires a time-ordered representation, which consists of a sequence of models ordered in time. The sequential ordering can correspond to one of several time scales. For example, the time-ordered models can be snapshots from molecular dynamics trajectories. They may also represent macromolecular assemblies at different stages of assembly or models at different cell cycle stages. Finally, the representation can be generalized to include an ensemble of models consistent with the input information. This section focuses on recent advances in multi-scale and multi-state representations for integrative models.

## Multi-scale modeling

Representations for integrative modeling are usually set manually and/or in an *ad hoc* manner. However, the choice of representation is important as it dictates the efficiency of exhaustive sampling, the accuracy with which input information is translated to spatial restraints, and the usefulness of the model for subsequent biological analysis. Recently, the concept of optimizing representation was introduced along with an objective criterion for optimizing representations [152]. The optimal representation was defined to be the highest resolution representation for which the sampling is exhaustive at a precision commensurate with the resolution of the representation. An incremental coarse-graining method for obtaining optimal coarse-grained representations for a system based on input information was described. Notably, the method produces non-uniformly coarse-grained representations, with regions associated with more (less) data represented at higher (lower) resolution. However, the method is computationally expensive since precise estimates of per-bead sampling precision are required for the incremental coarse-graining procedure. A new method, NestOR (Nested Sampling for Optimizing Representations) was developed to compare representations based on their fit to input data [153]. The fit to data is evaluated using the model evidence computed using the nested sampling algorithm. On a benchmark of assemblies, NestOR obtains optimal representations at a fraction of the time required to assess each representation *via* full-length production sampling. Methods such as NestOR can also be used to optimize other aspects of system representation such as the stoichiometry and number of rigid components. Further, an open question in multi-scale modeling is the optimal propagation of information across scales in simulation. Methods such as these will also be useful for meta-modeling, where the objective is to combine models at different scales (see Frontier 3).

## Multi-state modeling

Several methods have been developed to determine the optimal number of states or conformations that fit input information. Here, we discuss recent methods based on Bayesian inference and Bayesian model selection, since these explicitly account for uncertainty in the experimental data [6, 154]. Bayes factors were used to find the optimal set of states of proteins that describe SAS and NMR data (Potrzebowski et al. 2018). The model evidence term in the Bayes factor is maximized by a fast variational Bayes approach. The method works for restraints for which suitable analytical approximations of the posterior are available to accompany the variational Bayes method. This may not be feasible for all restraints. Bayes factors were also used to obtain the optimal number of chromatin states that explain ensemble Hi-C data [155]. The model evidence was calculated using the density of states method that involves sampling a series of annealed posteriors. Similarly, the method of BiCePs (Bayesian Inference of Conformational Populations) uses a Bayes factor-like term where the model evidence is computed using free-energy perturbation to reweigh conformations generated from theoretical models based on NMR and Hydrogen Deuterium Exchange (HDX) data [156, 157]. Both the density of states and free energy perturbation methods can be computationally expensive. Notably, these methods can be applied to refine an initial set of models generated, for example, from MD simulations.

Two multi-state approaches developed recently combine the maximum entropy approach with a Bayesian treatment of restraints from experimental data [6, 42, 46, 154, 158]. In the maximum entropy approach, a set of states (models) consistent with experimental data is obtained by minimally perturbing a set of initially generated models based on experimental data [154]. In contrast to previously described methods, these methods incorporate experimental data directly in sampling. Several parallel replicas are simulated under an MD force field with an additional restraint or energy term based on experimental data [42, 46, 158].

Modeling multiple states is particularly necessary for proteins that are dynamic and do not adopt a single conformation in solution, such as intrinsically disordered proteins (IDP). Several methods have been developed for disordered proteins. [159, 160] describe methods for Bayesian refinement of an initial ensemble using the maximum entropy approach. The methods reweigh initial IDP ensembles based on NMR, FRET, and SAXS data. Future goals for integrative multi-state modeling include developing methods for comparing, validating, and disseminating multi-state models [154].

## Improvements to sampling, analysis, and validation algorithms

Stochastic sampling algorithms such as Replica Exchange Markov Chain Monte Carlo (MCMC) are used for sampling the posterior distribution of models of macromolecular assemblies in Bayesian integrative modeling [11, 14, 154]. MCMC sampling methods involve several parameters such as move sizes of flexible beads and rigid bodies, which are usually tuned manually. A method, StoP (Stochastic Optimization of Parameters), was recently introduced to facilitate automatic tuning of these sampling parameters for the Integrative Modeling Platform (IMP) in a parallelized manner, based on a derivative-free, global, multi-objective optimization heuristic [161]. It allows for simultaneous parallel global optimization of several parameters, enhancing the efficiency of sampling the posterior distribution. However, it is not efficient at high dimensions of input parameters and is not guaranteed to find the optima in complex landscapes.

Several improvements were also made to the analysis and validation pipeline of integrative modeling. In particular, the first step after sampling earlier was to select good-scoring models based on thresholds on individual and/or total score terms [14]. These good-scoring models were then input into a set of four statistical tests to examine sampling exhaustiveness, determine the sampling precision, and identify model clusters [10]. The first step of selecting good-scoring models was altered to incorporate equilibration detection and density-based score clustering [11]. First, the equilibration of each score is determined for each independent run. Models that are produced after all scores equilibrate are subjected to clustering using Hierarchical DBSCAN (HDBSCAN), an assumption-free clustering method that can recognize arbitrary cluster shapes.

HDBSCAN is used to cluster models based on individual and total scores. This new, improved analysis provides a more unbiased way of choosing models for analysis.

Subsequently, models are subjected to sampling exhaustiveness tests [10, 11]. This includes distance-threshold-based clustering and quantifying the sampling and model precisions. Recent improvements to this part of the pipeline in IMP include the ability to account for ambiguity (*i.e.,* multiple copies of the same protein) in calculating the RMSD and precision of models and improvements to efficiency by parallelizing input loading and threshold-based clustering.

Model precision is usually defined at the level of the entire model. However, it would be useful to determine regions of low and high precision from integrative models. Regions of low precision can be used to determine where to gather the next set of experimental data to improve the model. High-precision regions can be used for subsequent in-depth biological analysis such as suggesting mutations and binding sites. A method to annotate regions of high- and low-precision on integrative models was developed, PrISM (precision for integrative structural models) [162]. First, a measure of bead-wise precision is computed from bead-wise 3D density grids. Beads are classified into low, medium, and high precision classes. Then, beads of similar precision that are also proximal in the integrative structure are clustered together to identify high- and low-precision regions. The method is versatile and can be applied to any set of integrative models.

Finally, the PDB-Dev database is a prototype archive of integrative (or hybrid) structures and associated data developed in the Worldwide Protein Data Bank (wwPDB) [20, 21]. It is expected that it will be integrated into the PDB in the near future. Validation pipelines are being developed for integrative models to be deposited to the PDB-Dev. The validation pipeline currently includes the assessment of models based on SAXS, chemical crosslinks, 3D EM data, and stereochemistry information. Ongoing and future work includes the development of validation methods for other types of experimental data, and the assessment of model uncertainty, both using Bayesian approaches.

# Frontiers in integrative modeling

Structural biology has been revolutionized over the past few years driven by advancements in cryo-electron microscopy (cryo-EM) and deep learning. Cryo-EM ignited a “resolution revolution” and allowed for structure determination at near-atomic resolutions [163]. Deep learning methods like AlphaFold [68] and RosettaFold (RF) [164] similarly allow for prediction of highly accurate atomic structures using sequence information alone. So, where does integrative modeling stand in the cryo-EM and AF2 era?

It is essential to acknowledge that not all challenges in structure determination are fully addressed by these techniques. As has also been addressed by [5], only 50% of the published EM maps have a resolution lower than 6 Å and several of these lack an accompanying PDB structure. Frequently, significant regions of constituent proteins are unresolved in published maps. Even the deep learning methods do not always yield accurate predictions. This may result due to the poor quality of the Multiple Sequence Alignment (MSA), presence of dynamic protein regions, and the bias of the trained model towards structures in the training dataset [165]. Often, individual domains of the proteins may be accurately predicted, but the relationship between domains might be incorrect [165]. Clearly, for such cases, integrative modeling is a preferred choice.

The need for an integrative structural biology approach has been underscored in recent advances in computation (AI-based structure prediction), experiments (cryo-ET), and theory (metamodeling). These are discussed below.

# Frontier 1. Integrating AI-based structure predictions with experiments

AlphaFold2 (AF2), developed by DeepMind, is a deep learning-based method for structure prediction using only sequence information [68, 166]. It has a wide-ranging utility as shown by the discovery of twenty-six new folds. Notably, an Alphafold database (AFDB) has been released that consists of approximately 200 million protein structures [167]. Further, it enabled proteome-wide modeling of protein-protein interactions (PPI) for *Escherichia coli* [166], *Saccharomyces cerevisiae* [168], and humans [169].

However, as also stated above, AlphaFold predictions are not reliable in case of a shallow MSA, the presence of intrinsically disordered regions (IDRs), or for predicting structures for proteins in large complexes [123, 146, 167]. Despite this, high-confidence predictions from AF2 can be utilized for structure prediction while predictions at lower confidence may still be used with appropriate validation. Overall, integrating deep learning-based methods such as AF2, with experimental data, such as chemical cross-links and cryo-EM/crystallographic maps, can improve structure determination (Figure 2) [146, 167].

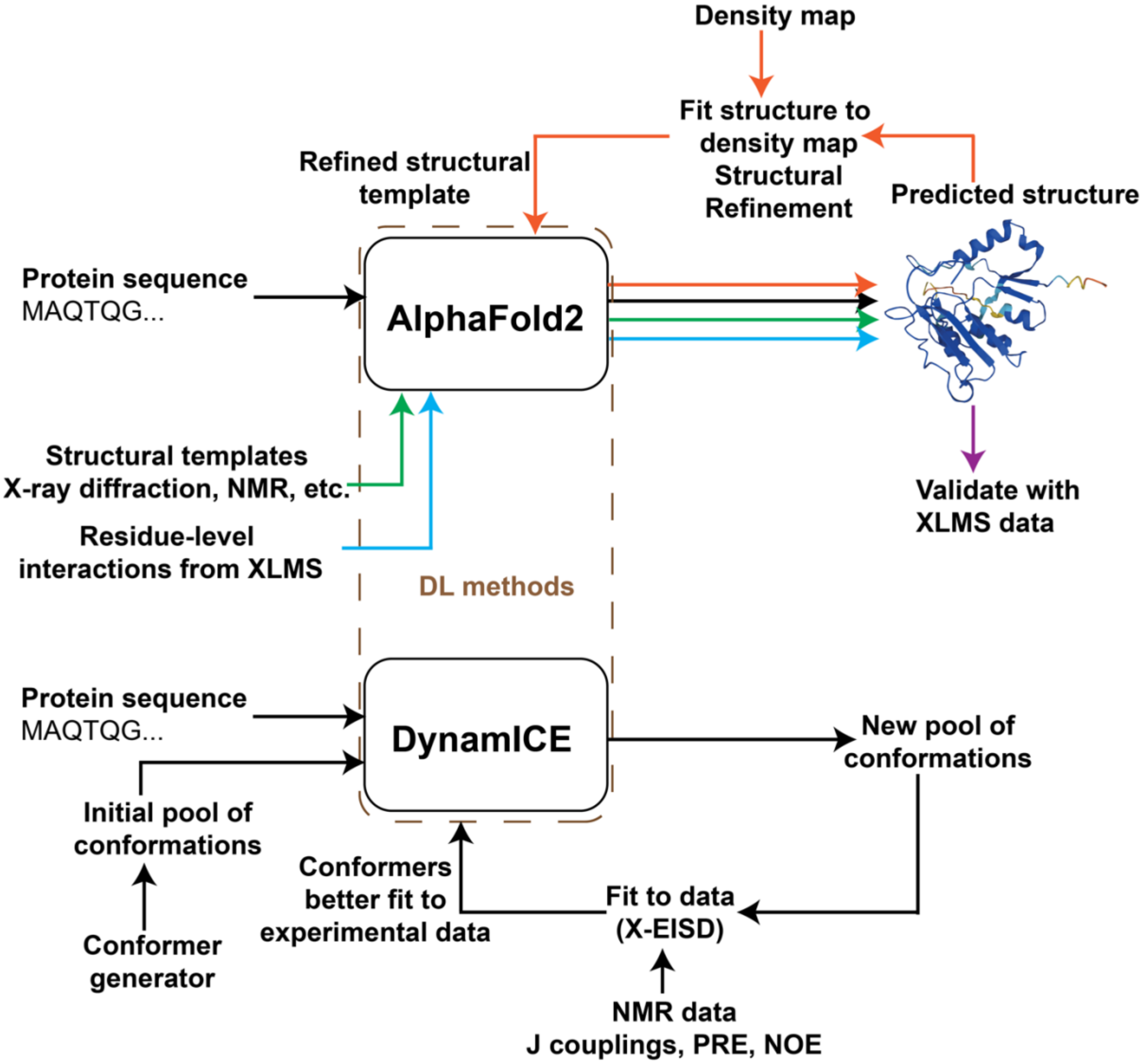

**Fig. 2 Frontiers in integrative modeling: combining AI-based structure predictions with experimental data** Schematic describing current methods that combine AI-based methods with experimental data for structure prediction. Experimental data such as XLMS (blue) and structural templates (green) can be used as additional inputs to improve the accuracy of Alphafold predictions. They can be used for validating AlphaFold predictions (purple). Additionally, AlphaFold predictions can be used to accelerate the process of structure determination using density maps from X-ray crystallography and/or cryo-EM (orange). Finally, deep generative modeling methods like DynamICE can be employed to refine structural ensembles to better fit the experimental data.

## AlphaFold predictions can be validated with crosslinks

Experimental data can be used to validate the predictions from AlphaFold2. Cross Linking mass spectrometry (XLMS) provides a means for assaying protein structures and protein-protein interactions (PPIs) under native conditions [146]. [146] state that crosslinking data can be used to gain insights on the structurally unresolved regions of a protein. They identified 28,910 unique cross-linked residue pairs of which 3,785 inter-protein, and 25,125 were intra-protein cross links. They used AlphaFold-multimer to obtain structures for 530 protein-protein interactions (PPIs) identified using the crosslinking data. They observe that AlphaFold-multimer could generate a well-modeled interface for 69% of these, significantly higher compared to 19% for randomly chosen protein pairs. From this, they conclude that AlphaFold-multimer can be used to identify PPIs but needs to be validated with experimental data like from XLMS.

On similar grouds, McCafferty C. L. et. al. (2023) [167] compare AF2-predicted structures of the 100 most crosslinked ciliary proteins identified by mass spectrometry from *T. thermophilia*. They observed a high concordance between the crosslinks and AF2 structure prediction along with some violations associated with domains of multi-domain proteins and domains which undergo major conformational changes.

O'Reilly F. J. et. al. (2023) [170] combined Alphafold-multimer predictions with XLMS data for proteome-wide PPI modeling. They identify 17 novel interactions without previous annotation and 130 interactions without close homologs in the PDB. Although the crosslinking information was not used in Alphafold2 model prediction, they observe a strong correlation between the i-pTM (interface-predicted Template Match) score and the satisfaction of heteromeric crosslinks.

## Chemical crosslinks can improve AlphaFold predictions

Besides its application in validating AlphaFold2 predictions, crosslinking data can also contribute to enhancing the accuracy of the predictions. AlphaLink integrates data from crosslinking photo-activable amino acids directly into the AlphaFold2 architecture [171]. Crosslinks bias the retrieval of evolutionary relationships by the Evoformer block in the AlphaFold architecture for updating the residue pair representations. Evolutionary information in turn compensates for the sparsity of crosslinks [68, 171]. This method allows for the investigation of multiple conformational states without relying on the manipulation of the mutiple sequence alignments (MSA). The authors also extended Alphalink for complex structure prediction by integrating cross linking data into AlphaFold multimer predictions [171].

## AlphaFold predictions can be improved by using atomic structures as templates

Another avenue for improving the Alphafold2 (AF2) predictions is to provide homologous PDB structures, obtained using X-ray crystallography, cryo-EM, and SAXS, as input templates. AF2 allows for incorporating the information from the templates directly into its architecture.

Mirabello C. *et*. al. combine complementary information from multiple sequence alignments (MSAs) and structural templates for improving AlphaFold-multimer predictions [172]. AlphaFold-multimer does not use any information about the inter-chain interactions in quaternary templates. Mirabello C. and co-workers integrate this information from templates containing quaternary structures in the AlphaFold architecture. This modification removes the mask on inter-chain distances in the Alphafold architecture and importantly, does not require retraining. The integrated templates could either be a PDB structure or a draft of an experimental structure. Structural alignment is used to select templates for building a homology model for each chain. The information derived from the templates and MSAs is then used for structure prediction by Alphafold. Further, they suggest using Alphafold for structural inpainting or filling in the missing residues in the PDB structure. They use the partial available structure as a template and the sequence of the protein as the input for AlphaFold and allowed it to predict the complete structure including the missing residues.

## AlphaFold predictions improve the efficiency of X-ray and cryo-EM structure determination

Alphafold predictions can be used to accelerate X-ray and cryo-EM-based structure determination. Instead of using the atomic structures from these experiments directly as templates for Alphafold2, information from electron density maps can be incorporated into Alphafold2 for obtaining high-resolution structures much faster than the conventional methods.

[165] developed a tool for accelerating structure determination from X-ray data by iteratively refining AlphaFold-predicted structures with electron density maps. Starting from an initial AF2 prediction, the model building proceeds by cycles of fitting the AF2-predicted structure onto the

electron density map, refining, and using the refined structure as a template for AF2. This method could successfully predict at least partially correct structures for 187 out of 215 datasets. This and other similar methods rely on the fact that the template used for rebuilding an AlphaFold model using a density map contains information from that density map [173].

## Integrative modeling using generative AI

Generative modeling with experimental data represents another compelling frontier in the realm of structural biology. Deep generative models are currently employed in protein design, tasked with producing an amino acid sequence based on a provided protein structure. RFdiffusion (RosettaFold Diffusion), an open- source tool developed by the Baker Lab is one example [174].

Generative modeling has also been used for generating an ensemble of conformations for intrinsically disordered proteins (IDPs) in particular. Intrinsically disordered proteins (IDPs) are a class of proteins which lack a well-defined structure in the monomeric state and exist in an ensemble of conformations. Characterization of IDPs poses a significant challenge both experimentally and computationally [175, 176]. DynamICE (dynamic IDP creator with experimental restraints) is designed to generate a set of conformations for IDPs that are consistent with NMR data including NMR J couplings, paramagnetic resonance enhancements (PRE), and nuclear overhauser effects (NOE) [176]. An initial pool of IDP conformations is generated by IDP-specific sampling methods such as fragment-based methods. Given an input sequence and torsion angles of the first residue in the sequence from conformations in the current pool, DynamICE uses a generative recurrent neural network (GRNN) to learn the probability of the successive residue torsion angles and generate a new pool of conformers. A reinforcement learning (RL) step in the GRNN assesses the fit of the new pool with the experimental data using extended Experimental Inferential Structure Determination (X-EISD) Bayesian approach [160]. The new pool forms the input to the GRNN in the next step, and the process is repeated until convergence. This generates a pool of conformations that better fit the experimental data compared to the input pool. DynamICE is comparatively less expensive than MD simulations for conformational sampling and considers the experimental errors and uncertainties objectively, in a Bayesian manner. However, its use of a torsion-based representation makes it less suitable for experimental data that provides distance restraints like Nuclear Overhauser Effect (NOE), Paramagnetic Relaxation effect (PRE).

# Frontier 2. Using data from cryo electron-tomography for integrative modeling

Cryo-electron tomography (cryo-ET) is an imaging technique that enables 3D visualization of the native cellular environment at sub-nanometer resolution, providing unpreceded insights into the molecular organization of cells [177]. The acquired images can then be used to reconstruct 3D volumes at nanometer resolution, allowing for the structural characterization of macromolecules while preserving their cellular context [178]. Structural characterization of macromolecules using cryo-ET involves localizing the particles of interest in a tomogram, subtomogram alignment, classification, and averaging, and subsequent post-processing [179]. However, the compositional heterogeneity, the missing wedge, the low signal-to-noise ratio in the data, and the crowded nature/distribution of particles in tomograms make the task challenging [179, 180]. Owing to the recent advances in data collection methods, it has now become possible to generate large tomographic datasets within a day [181].

Cryo-ET is a crucial tool in the field of visual proteomics, wherein, the goal is to generate molecular atlases, *i.e.*, localization and identification of macromolecular assemblies within a cell [182]. Localization of an assembly refers to the annotation of all the locations in the 3D tomogram corresponding to densities of the assembly. Identification is the assignment of an unknown density to a macromolecular assembly. These atlases can be used for *in situ* structural characterization of macromolecular assemblies. The localization and identification of assemblies requires cryo-ET data along with templates generated from various structure determination techniques and AI-based structure predictions. The *in situ* structural characterization requires subtomogram averaging (STA) to build 3D maps which can then be utilized along with data from other experimental techniques such as fluorescence microscopy (for localizations), cross-linking coupled with mass spectrometry (XLMS, for interactions), and NMR, X-ray crystallography, single particle analysis with cryo-EM and AI-based structure predictions [182, 183]. Therefore, localization, identification and *in situ* structural characterization are inherently integrative. The methods for structural characterization of macromolecular assemblies using cryo-ET data have only recently been developed. Some of these approaches are described below.

## Localization and Identification of Macromolecular Assemblies in Tomograms

The low signal-to-noise ratio in the tomograms makes localization and identification of assemblies (referred to as particles) difficult [178]. Manual particle annotation is possible for some large macromolecular assemblies, however, it is a laborious and time-consuming task and is not applicable to most macromolecular assemblies. Moreover, it is a rate-limiting step in the high-throughput characterization of these assemblies [177]. As a means of promoting the development of new methods for extracting information from tomograms, [178] create novel datasets as a SHREC (Shape Retrieval Challenge) track as part of the 3D Object Retrieval Workshop Series and benchmark several particle identification and localization tools [178, 184, 185].

### Template matching

Template matching is one of the most commonly used methods to localize particles corresponding to a target molecular species in tomograms. In this method, a template is generated using a low-pass filter on the structural model of the target. The densities within the tomogram that have high cross-correlation with this template are identified as the particles corresponding to the target species [186]. However, for template matching, the structure (or the shape) of the macromolecule to be characterized needs to be known, making it unusable for *de novo* identification and localization of macromolecules in the tomograms. Moreover, since cross-correlation between the template and the densities need to be computed at all points and in all orientations, it is computationally expensive [177]. While this approach performs well for localizing large assemblies, it performs poorly on small complexes and in crowded environments. Also, it fails to distinguish complexes that are structurally similar, for example, multiple states of a given complex [180].

### Deep learning-based methods

Owing to the advancements in the field of computer vision, several deep learning-based methods have been developed to automate the localization and identification of macromolecular assemblies in tomograms [177, 180, 181, 187–192]. These methods are significantly faster than template matching and better suited for multi-class problems.

**Deepfinder** [180] have proposed a deep learning-based semantic segmentation method, DeepFinder, for the simultaneous identification and localization of several macromolecular species from tomograms. DeepFinder utilizes a 3D convolutional neural network (CNN), based on the U-net architecture, to perform multi-class semantic segmentation of input tomograms. It was the best-performing method in SHREC 2019 and the second best in SHREC 2020 challenges [178, 180, 185]. The method requires a relatively small training dataset and performs better than template matching [178, 180, 185]. It is trained on expert annotated tomograms for several different macromolecular species and can be used for the identification and localization of particles corresponding to these macromolecular species in new tomograms. The training dataset can be expanded to include more macromolecular species by adding segmented tomograms for these target species. Similar to template matching, a limitation of DeepFinder is its inability to identify and localize small macromolecular species in the tomograms. Moreover, it produces a large number of false positives and also detects poorly resolved particles from the tomograms, which affects the resolution of downstream structural characterization [180].

**DeePiCt** DeePiCt (Deep Picker in Context), another similar supervised learning method, provides an added level of complexity to this approach, wherein, it allows for the analysis of particles within their cellular context. It uses a 2D CNN to localize cellular organellar structures, such as membranes, and a 3D CNN with tunable hyperparameters to identify and localize macromolecular assemblies. The outputs from the two networks are then integrated in one of three ways - intersection, colocalization, and contact. Such an approach makes the structural comparison of the particles of the same assembly in different spatial contexts possible. Moreover, this approach reduces the number of false positives, which in turn improves the quality of output from the downstream STA [189].

**SPHIRE-crYOLO** [190] proposed a supervised learning approach (crYOLO) that is based on the YOLO (You Only Look Once) object detection method. It was originally developed for use with single-particle analysis datasets and the support for tomogram datasets was added later [179]. Instead of passing a sliding window across the image, which is the basis for CNNs, crYOLO divides the image into grids and passes the entire image at once through the neural network. The network then identifies whether the grid cell contains the center of the bounding box for a target particle. If a grid cell contains the bounding box center, the network then identifies the relative position of the center of the particle within the grid cell. The YOLO approach makes crYOLO significantly faster than traditional CNNs while retaining their accuracy.

**MemBrain** Owing to their small size and buried nature, the identification and localization of membrane-embedded proteins is a challenging task. [177] have developed a deep learning-aided object detection pipeline for the identification and localization of membrane-embedded proteins. Their approach takes segmented membranes from raw tomograms as inputs. Their method starts by uniformly drawing samples along the segmented membrane and identifying the membrane normal vectors. Subvolumes are then extracted around each sampled point and aligned such that the membrane lies parallel to the x-y plane.

Prior to training, the subvolumes are labeled corresponding to the Mahalanobis distance of the center of the subvolume from the center of the nearest protein particle. A CNN is trained on the aligned subvolumes and their corresponding distance labels. It then predicts distances for the target membrane, which are combined to generate the distance map for the target membrane segmentation. The distance map is clustered using hierarchical Mean Shift clustering to predict the particle centers. The orientations of the particles are then determined relying on the assumption that the converged cluster points resemble the rough shape of the protein complex. The requirement of a small training set (one annotated membrane) and the ability to estimate the particle orientations are some of the key advantages of this method.

**Deep metric learning** [188] propose a deep metric learning approach. Their model learns the generalized representation of a macromolecular shape. A CNN-based architecture is used to learn an embedding on sub-tomograms with known macromolecules, such that particles of a given macromolecule are close together and unrelated macromolecules are far apart in the embedding space. This architecture allows the model to place the particles corresponding to novel macromolecules based on their similarity with the known macromolecules in the embedding space. However, this approach has some limitations. Currently, it is not implemented for use with membrane-bound macromolecules. Moreover, the clustering in the embedding space is dependent on the copy number of the macromolecule in the tomogram. Finally, this method cannot distinguish between different structural configurations of the same macromolecule.

**Few-shot supervised learning approaches** Manually annotating target proteins and macromolecular assemblies in tomographic datasets for training the supervised learning methods is time-consuming and often infeasible as the structures for several proteins are still unknown. Towards this end, few-shot supervised learning approaches have been developed, which require very few labeled examples of the target class. In few-shot learning, a model is pre-trained on a training dataset comprising a large number of labeled samples of known structures, which provide the prior knowledge. Then, a few labeled samples for the target macromolecule (support set) are used to adapt the model to predict the unlabeled samples (query examples). The key advantage of this approach is that a low number of annotated particles per target class is sufficient for training and thus it retains the benefits of a supervised learning method while reducing the need for training data [187].

[187, 192] propose few-shot learning approaches for particle identification from tomograms. The approach proposed by [187] used a modification of the state-of-the-art prototypical network (ProtoNet), ProtoNet3D, to generate a task-agnostic embedding which is then used as input for a transformer to generate a task-specific embedding. They call it the ProtoNet-CE (Protonet with Combined Embedding) model. It uses both these embeddings for performing particle classification using a nearest-neighbor classifier. Another few-shot approach developed by ([187, 192]) additionally corrects for the fact that the tomograms in the training and test sets may vary in their intensities. For example, the training set (source domain) may be obtained from simulations while the test set (target domain) is sourced from real data. Their approach, termed Few-Shot Fine-Tuning domain adaptation, consists of three networks, an encoder, which extracts a feature vector from each subtomogram, a classifier, that provides the structural class of each feature vector, and a discriminator that provides the domain of each feature vector. First, the encoder and classifier are trained on the source domain. Next, keeping the encoder and classifier networks fixed, the discriminator is trained to identify the domain of a subtomogram. Finally, keeping the discriminator fixed, the encoder and classifier are fine-tuned using labeled data from both domains. This approach requires only a few (less than ten) labeled examples in the target domain.

Supervised learning approaches, while good at feature extraction and robust against noise in the dataset, leave the structural characterization workflow limited to particles with known shape [187, 188]. In order to identify a novel target, these models need to be retrained on a new dataset comprising tomograms and the corresponding annotations for the target. While the few-shot supervised approaches above reduce the requirement of training samples, they still limit the high-throughput characterization of macromolecules and their assemblies observed in cryo-ET datasets [187, 188]. To facilitate the detection of macromolecules of unknown structure, unsupervised learning approaches have been developed [181, 191]. Some of these are described below.

**MPP** [191] developed the Multi-Pattern Pursuit (MPP) framework for identifying frequently occurring structural patterns in tomograms. They define a structural pattern as a set of subtomograms that are likely to contain particles of similar structure, and that when averaged, produce the density map of the particle. Given a set of tomograms, their workflow starts with template-free particle picking based on the Difference of Gaussian (DoG) filtering. The extracted sub-tomograms that contain particles of similar size and shape are then normalized by translation and pose and clustered. These subtomograms are passed on to the MPP framework which first iteratively searches for structural patterns based on the current rigid transformation of each sub-tomograms in a cluster using an imputation-based dimension reduction and clustering approach. These patterns are subsequently refined by alignment and redundant pattern removal aiming to maximize the quality of the extracted patterns [191].

**DISCA** With the aim of performing high-throughput systematic comprehensive analysis of cryo-ET datasets, [181] proposed an unsupervised deep learning approach named DISCA (Deep Iterative Subtomogram Clustering Approach). It starts with template-free particle picking followed by subtomogram extraction. Utilizing a generalized Expectation Maximization (EM) framework, DISCA iteratively clusters these subtomograms based on their 3D structural features extracted by CNNs. The subtomogram subsets can then be aligned, averaged and re-embedded in the tomogram for their *in situ* structural characterization [181].

## *In situ* structural characterization of assemblies using cryo-ET data with AI-based predictions

Current workflows for the structure determination of assemblies require data from several experiments - fluorescence imaging to obtain its localization within the cell, mass spectrometry to identify component proteins, followed by *in vitro* structural characterization. This workflow is time-consuming and the modeled structures might differ from the native configuration of the target protein or macromolecular assembly [101].

[101] propose an alternate method for assigning proteins to unidentified densities in the 3D reconstruction maps obtained from cryo-tomography data utilizing predictions from AlphaFold2. Their results were further validated by biochemical assays and mass spectrometry experiments. In the generated 3D maps of mouse and human sperm axonemes, the microtubule doublets were identified by comparison with previously characterized cryo-EM single particle analysis maps of bovine tracheal ciliary doublets. Similarly, the densities corresponding to 29 Microtubule Inner Proteins (MIPs) were identified and were used to build a pseudo-atomic model. The densities that were unaccounted for were computationally isolated. These densities that were unaccounted for, were then matched the entries in the AlphaFold2 mouse proteome library and ranked by their cross-correlation scores [101].

Their integrative workflow thus combines high-resolution cryo-tomography data with predictions from AlphaFold2 and previously published structures to identify and structurally characterize the densities that were otherwise unaccounted for. Similar approaches can be utilized to identify novel macromolecular assemblies within a cell that are recalcitrant to structure determination by conventional methods. Moreover, the cellular context that is intrinsic to such an approach may provide further insights into the functions and interactions of these macromolecules in their native environment.

# Frontier 3. Spatio-temporal models of cells

Recent efforts to compute spatio-temporal models of entire cells have extended the scale of structural biology from single protein structures, complexes, and assemblies to whole cells [1]. Here, we discuss three recent approaches to whole-cell modeling.

## Static structural whole-cell models

Static, atomically-detailed, mesoscale (10-100 nm) structural models of heterogeneous aspects of the cell have been modeled by cellPACK [193]. It relies on AutoPACK, an algorithm for packing geometrical objects in a global grid representing cell, under geometric constraints [193]. Atomic structures, for instance, are represented as spherical beads and packed as rigid bodies.

Recently, the structure of the *Mycoplasma genitalium* (MG) cell was determined based on genomics and proteomics information, known atomic structures and homology models of proteins, nucleoid models, and previous systems biology simulations of the MG cell that informed ribosome status, state of translation of the genome, protein concentrations, and DNA-binding protein occupancies of each nucleotide [194]. This work is a starting point for correlating changes in structure to cell cycle state. cellPACK was also used to test hypotheses on the distribution of envelope proteins in HIV virions. Data from fluorescence microscopy and cryo-electron microscopy differed in their distributions of the envelop protein, with one of them being symmetric and the other being asymmetric. The cellPACK model was used to assess the consistency of each of the above sources with other input information [193].

In contrast to integrative modeling *via* IMP, for instance, one can model larger components such as membranes, filamentous actin, and so on. Other useful features of cellPACK include its open-source software, modular design, the ability to mix-and-match constraints for packing, support for iterative refinement of models and community-sourced information, and support for multi-scale modeling with a versatile set of geometric primitives. Apart from being starting points for dynamics simulations, these models are also meant to be easy to generate and use for pedagogical and outreach purposes. The disadvantages are that the approach is semi-quantitative, computes only static structures, supports a small number of input experiment types, and does not account for uncertainty in the input information.

## Molecular dynamical and fully dynamical whole-cell models

A completely quantitative approach has been adopted to determine fully dynamical whole-cell models. A minimal bacterial cell, JCVI-Syn3A, was simulated using fully dynamical kinetic models of a network of chemical reactions [195, 196]. A hybrid multi-scale approach involving a combination of reaction-diffusion equations, chemical master equations, and ordinary differential equations was used. All molecules, including proteins, lipids, and DNA were represented as spheres. Simulations were based on information on the cell shape and ribosome distributions from cryo-ET, chromatin organization from Chip-Seq, protein and membrane composition from proteomics and lipidomics, and concentrations of molecules from transcriptomics and metabolomics. Kinetic parameters and molecular concentrations were inferred using a Bayesian approach from related cells. The model was validated by agreement with various experiments, including surface area doubling time and mRNA half-life measurements. Time-dependent concentrations of molecules were obtained from the model as a function of cell cycle. The model identified active transport as the cellular process that consumes the most energy. Several emergent properties were also identified. In particular, the effect of reduced metabolism on slowing down of DNA replication and transcription was quantified. Despite this advance, several challenges remain. For one thing, input information is incomplete. The functions and kinetics of all proteins in the proteome are not available, even for a minimal cell. Due to this, for example, some complexes such as the transcription machinery, are represented in their inactive state. Concentrations of molecules in different parts of the cytosol, as well as the architecture of the cell is not available. Several omics studies are performed on populations and not on single cells and data is not available for all stages of the cell cycle. On the computational method development front, there is a need to develop more realistic models for interactions between molecules, instead of the current simplified models such as excluded volume.

Molecular dynamics simulations of minimal cells were also attempted. However, even coarse-grained models of such cells exceed our current capabilities. They are challenging for various reasons, including the technical difficulties of storing in memory and simulating hundreds of millions of particles for reasonably long time scales, the non-availability of coarse-grained models of several complexes, and the need to fine-tune the balance between ionic solvents and lipids [197].

## Metamodeling

A third approach encompasses both the static and dynamic models above. Several models of the cell already exist at various scales, such as models of kinetics of reactions in the cell, models of pathways, and structural models of proteins and complexes. One way to combine these models is to integrate data from multiple sources and use a sophisticated modeling method to compute a single multi-scale model. However, in an alternative approach, a recent advance aims to integrate the computed models themselves [1, 198]. The theoretical framework to combine models is known as Bayesian metamodeling. Probabilistic graphical models known as surrogate models encode the uncertainty in each model. The relationship between individually computed models is used to construct couplers which are conditional probability distribution functions. Finally, the joint probability distribution of the surrogate models and couplers is determined and used to harmonize the models with respect to any new information through back-propagation. Several advantages of this approach are apparent. First, it facilitates a distributed approach to model building. Each type of model requires significant domain expertise and it would be impossible for one or even a few research labs to gather this expertise to compute a whole-cell model by themselves. The divide-and-conquer approach allows for models on specific aspects of cells to be computed by groups focused on those aspects. These models can then build on each other. It facilitates collaboration between groups. Second, it provides a framework to detect and resolve conflicts between models, as each model is coupled to, and harmonized with other models. Third, the uncertainty of each model is explicitly incorporated. Finally, the approach is more efficient than combining all data and computing a multi-scale model in one go [1]. Proofs-of-concept were shown on the model of glucose-stimulated insulin secretion by pancreatic β-cells and dynamics of cellular networks in an islet [198, 199]. However, models may not always be coupled usefully to yield predictive insight. Also, it is non-trivial to organize collaborations among diverse groups. Future improvements include development of better coupling functions, loss functions, optimization algorithms, and validation methods for automated metamodeling.

# Conclusion

Integrative structures are now archived by the wwPDB, in a manner similar to X-ray or NMR structures, underscoring their importance [21, 200]. Recent advances in experiments (cryo-ET), computation (AI-based structure prediction), and theory (metamodeling) highlight the need for adopting an integrative structural biology approach. Known atomic structures from X-ray or single-particle cryo-EM are fit into maps from electron cryo-tomography, an emerging experimental technique for characterizing structures *in situ*. Structures predicted by deep-learning based methods such as Alphafold are now used in conjunction with other experimental data, such as chemical crosslinks and cryo-EM maps, to build more complete and accurate models of large macromolecular assemblies. Finally, metamodeling allows us to integrate multiple independently computed models to build spatio-temporal models of entire cells. Taken together, these advances carry the promise of a significant increase in the accuracy, resolution, scale, precision, and efficiency of structures determined, thereby making structure-based hypotheses and inferences about function accessible to any biologist.

# Acknowledgements

We thank lab members Muskaan Jindal, Arijit Das, and Sakshi Shigvan, for useful comments on the manuscript. This work has been supported by the Department of Atomic Energy (DAE) TIFR grant RTI 4006 and Department of Science and Technology (DST) SERB grant SPG/2020/000475 from the Government of India.

# References

1. Sali, A.: From integrative structural biology to cell biology. Journal of Biological Chemistry. 296, 100743 (2021). https://doi.org/10.1016/j.jbc.2021.100743
2. Carugo, O., Djinović-Carugo, K.: Structural biology: A golden era. PLOS Biology. 21, e3002187 (2023). https://doi.org/10.1371/journal.pbio.3002187
3. Kendrew, J.C., Bodo, G., Dintzis, H.M., Parrish, R.G., Wyckoff, H., Phillips, D.C.: A three-dimensional model of the myoglobin molecule obtained by x-ray analysis. Nature. 181, 662–666 (1958). https://doi.org/10.1038/181662a0
4. RCSB PDB: Homepage, https://www.rcsb.org/
5. Braitbard, M., Schneidman-Duhovny, D., Kalisman, N.: Integrative Structure Modeling: Overview and Assessment. Annual Review of Biochemistry. 88, 113–135 (2019). https://doi.org/10.1146/annurev-biochem-013118-111429
6. Habeck, M.: Bayesian methods in integrative structure modeling. Biological Chemistry. 404, 741–754 (2023). https://doi.org/10.1515/hsz-2023-0145

7. Koukos, P.I., Bonvin, A.M.J.J.: Integrative Modelling of Biomolecular Complexes. Journal of Molecular Biology. 432, 2861–2881 (2020). https://doi.org/10.1016/j.jmb.2019.11.009
8. Rantos, V., Karius, K., Kosinski, J.: Integrative structural modeling of macromolecular complexes using Assembline. Nat Protoc. 17, 152–176 (2022). https://doi.org/10.1038/s41596-021-00640-z
9. Rout, M.P., Sali, A.: Principles for Integrative Structural Biology Studies. Cell. 177, 1384–1403 (2019). https://doi.org/10.1016/j.cell.2019.05.016
10. Viswanath, S., Chemmama, I.E., Cimermancic, P., Sali, A.: Assessing Exhaustiveness of Stochastic Sampling for Integrative Modeling of Macromolecular Structures. Biophysical Journal. 113, 2344–2353 (2017). https://doi.org/10.1016/j.bpj.2017.10.005
11. Saltzberg, D.J., Viswanath, S., Echeverria, I., Chemmama, I.E., Webb, B., Sali, A.: Using Integrative Modeling Platform to compute, validate, and archive a model of a protein complex structure. Protein Sci. 30, 250–261 (2021). https://doi.org/10.1002/pro.3995
12. Lasker, K., Phillips, J.L., Russel, D., Velázquez-Muriel, J., Schneidman-Duhovny, D., Tjioe, E., Webb, B., Schlessinger, A., Sali, A.: Integrative structure modeling of macromolecular assemblies from proteomics data. Mol Cell Proteomics. 9, 1689–1702 (2010). https://doi.org/10.1074/mcp.R110.000067
13. Alber, F., Dokudovskaya, S., Veenhoff, L.M., Zhang, W., Kipper, J., Devos, D., Suprapto, A., Karni-Schmidt, O., Williams, R., Chait, B.T., Rout, M.P., Sali, A.: Determining the architectures of macromolecular assemblies. Nature. 450, 683–694 (2007). https://doi.org/10.1038/nature06404
14. Saltzberg, D., Greenberg, C.H., Viswanath, S., Chemmama, I., Webb, B., Pellarin, R., Echeverria, I., Sali, A.: Modeling Biological Complexes Using Integrative Modeling Platform. Methods Mol Biol. 2022, 353–377 (2019). https://doi.org/10.1007/978-1-4939-9608-7_15
15. Bonomi, M., Hanot, S., Greenberg, C.H., Sali, A., Nilges, M., Vendruscolo, M., Pellarin, R.: Bayesian Weighing of Electron Cryo-Microscopy Data for Integrative Structural Modeling. Structure. 27, 175-188.e6 (2019). https://doi.org/10.1016/j.str.2018.09.011
16. Bonomi, M., Pellarin, R., Kim, S.J., Russel, D., Sundin, B.A., Riffle, M., Jaschob, D., Ramsden, R., Davis, T.N., Muller, E.G.D., Sali, A.: Determining Protein Complex Structures Based on a Bayesian Model of in Vivo Förster Resonance Energy Transfer (FRET) Data*. Molecular & Cellular Proteomics. 13, 2812–2823 (2014). https://doi.org/10.1074/mcp.M114.040824
17. Saltzberg, D.J., Hepburn, M., Pilla, K.B., Schriemer, D.C., Lees-Miller, S.P., Blundell, T.L., Sali, A.: SSEThread: Integrative threading of the DNA-PKcs sequence based on data from chemical cross-linking and hydrogen deuterium exchange. Progress in Biophysics and Molecular Biology. 147, 92–102 (2019). https://doi.org/10.1016/j.pbiomolbio.2019.09.003
18. Schneidman-Duhovny, D., Hammel, M., Tainer, J.A., Sali, A.: FoXS, FoXSDock and MultiFoXS: Single-state and multi-state structural modeling of proteins and their complexes based on SAXS profiles. Nucleic Acids Research. 44, W424–W429 (2016). https://doi.org/10.1093/nar/gkw389
19. Shi, Y., Fernandez-Martinez, J., Tjioe, E., Pellarin, R., Kim, S.J., Williams, R., Schneidman-Duhovny, D., Sali, A., Rout, M.P., Chait, B.T.: Structural characterization by cross-linking reveals the detailed architecture of a coatomer-related heptameric module from the nuclear pore complex. Mol Cell Proteomics. 13, 2927–2943 (2014). https://doi.org/10.1074/mcp.M114.041673
20. Sali, A., Berman, H.M., Schwede, T., Trewhella, J., Kleywegt, G., Burley, S.K., Markley, J., Nakamura, H., Adams, P., Bonvin, A.M.J.J., Chiu, W., Peraro, M.D., Di Maio, F., Ferrin, T.E., Grünewald, K., Gutmanas, A., Henderson, R., Hummer, G., Iwasaki, K., Johnson, G., Lawson, C.L., Meiler, J., Marti-Renom, M.A., Montelione, G.T., Nilges, M., Nussinov, R., Patwardhan, A., Rappsilber, J., Read, R.J., Saibil, H., Schröder, G.F., Schwieters, C.D.,

Seidel, C.A.M., Svergun, D., Topf, M., Ulrich, E.L., Velankar, S., Westbrook, J.D.: Outcome of the First wwPDB Hybrid/Integrative Methods Task Force Workshop. Structure. 23, 1156–1167 (2015). https://doi.org/10.1016/j.str.2015.05.013

21. Vallat, B., Webb, B., Fayazi, M., Voinea, S., Tangmunarunkit, H., Ganesan, S.J., Lawson, C.L., Westbrook, J.D., Kesselman, C., Sali, A., Berman, H.M.: New system for archiving integrative structures. Acta Crystallogr D Struct Biol. 77, 1486–1496 (2021). https://doi.org/10.1107/S2059798321010871
22. Dominguez, C., Boelens, R., Bonvin, A.M.J.J.: HADDOCK: a protein-protein docking approach based on biochemical or biophysical information. J Am Chem Soc. 125, 1731–1737 (2003). https://doi.org/10.1021/ja026939x
23. Karaca, E., Bonvin, A.M.J.J.: On the usefulness of ion-mobility mass spectrometry and SAXS data in scoring docking decoys. Acta Cryst D. 69, 683–694 (2013). https://doi.org/10.1107/S0907444913007063
24. Emsley, P., Lohkamp, B., Scott, W.G., Cowtan, K.: Features and development of Coot. Acta Cryst D. 66, 486–501 (2010). https://doi.org/10.1107/S0907444910007493
25. Emsley, P., Cowtan, K.: Coot: model-building tools for molecular graphics. Acta Cryst D. 60, 2126–2132 (2004). https://doi.org/10.1107/S0907444904019158
26. Adams, P.D., Afonine, P.V., Bunkóczi, G., Chen, V.B., Davis, I.W., Echols, N., Headd, J.J., Hung, L.-W., Kapral, G.J., Grosse-Kunstleve, R.W., McCoy, A.J., Moriarty, N.W., Oeffner, R., Read, R.J., Richardson, D.C., Richardson, J.S., Terwilliger, T.C., Zwart, P.H.: PHENIX: a comprehensive Python-based system for macromolecular structure solution. Acta Crystallogr D Biol Crystallogr. 66, 213–221 (2010). https://doi.org/10.1107/S0907444909052925
27. Pfab, J., Phan, N.M., Si, D.: DeepTracer for fast de novo cryo-EM protein structure modeling and special studies on CoV-related complexes. Proceedings of the National Academy of Sciences. 118, e2017525118 (2021). https://doi.org/10.1073/pnas.2017525118
28. Chang, L., Wang, F., Connolly, K., Meng, H., Su, Z., Cvirkaite-Krupovic, V., Krupovic, M., Egelman, E.H., Si, D.: DeepTracer-ID: De novo protein identification from cryo-EM maps. Biophys J. 121, 2840–2848 (2022). https://doi.org/10.1016/j.bpj.2022.06.025
29. Orzechowski, M., Tama, F.: Flexible Fitting of High-Resolution X-Ray Structures into Cryoelectron Microscopy Maps Using Biased Molecular Dynamics Simulations. Biophys J. 95, 5692–5705 (2008). https://doi.org/10.1529/biophysj.108.139451
30. Trabuco, L.G., Villa, E., Schreiner, E., Harrison, C.B., Schulten, K.: Molecular Dynamics Flexible Fitting: A practical guide to combine cryo-electron microscopy and X-ray crystallography. Methods. 49, 174–180 (2009). https://doi.org/10.1016/j.ymeth.2009.04.005
31. Trabuco, L.G., Villa, E., Mitra, K., Frank, J., Schulten, K.: Flexible Fitting of Atomic Structures into Electron Microscopy Maps Using Molecular Dynamics. Structure. 16, 673–683 (2008). https://doi.org/10.1016/j.str.2008.03.005
32. Beton, J.G., Mulvaney, T., Cragnolini, T., Topf, M.: Cryo-EM structure and B-factor refinement with ensemble representation. Nat Commun. 15, 444 (2024). https://doi.org/10.1038/s41467-023-44593-1
33. Terashi, G., Wang, X., Maddhuri Venkata Subramaniya, S.R., Tesmer, J.J.G., Kihara, D.: Residue-wise local quality estimation for protein models from cryo-EM maps. Nat Methods. 19, 1116–1125 (2022). https://doi.org/10.1038/s41592-022-01574-4
34. Han, X., Terashi, G., Christoffer, C., Chen, S., Kihara, D.: VESPER: global and local cryo-EM map alignment using local density vectors. Nat Commun. 12, 2090 (2021). https://doi.org/10.1038/s41467-021-22401-y

35. Rotkiewicz, P., Skolnick, J.: Fast procedure for reconstruction of full-atom protein models from reduced representations. J Comput Chem. 29, 1460–1465 (2008). https://doi.org/10.1002/jcc.20906
36. Song, Y., DiMaio, F., Wang, R.Y.-R., Kim, D., Miles, C., Brunette, T., Thompson, J., Baker, D.: High-Resolution Comparative Modeling with RosettaCM. Structure. 21, 1735–1742 (2013). https://doi.org/10.1016/j.str.2013.08.005
37. Shekhar, M., Terashi, G., Gupta, C., Sarkar, D., Debussche, G., Sisco, N.J., Nguyen, J., Mondal, A., Vant, J., Fromme, P., Van Horn, W.D., Tajkhorshid, E., Kihara, D., Dill, K., Perez, A., Singharoy, A.: CryoFold: Determining protein structures and data-guided ensembles from cryo-EM density maps. Matter. 4, 3195–3216 (2021). https://doi.org/10.1016/j.matt.2021.09.004
38. Terashi, G., Kihara, D.: De novo main-chain modeling for EM maps using MAINMAST. Nat Commun. 9, 1618 (2018). https://doi.org/10.1038/s41467-018-04053-7
39. Singharoy, A., Teo, I., McGreevy, R., Stone, J.E., Zhao, J., Schulten, K.: Molecular dynamics-based refinement and validation for sub-5 Å cryo-electron microscopy maps. eLife. 5, e16105 (2016). https://doi.org/10.7554/eLife.16105
40. MacCallum, J.L., Perez, A., Dill, K.A.: Determining protein structures by combining semireliable data with atomistic physical models by Bayesian inference. Proc. Natl. Acad. Sci. U.S.A. 112, 6985–6990 (2015). https://doi.org/10.1073/pnas.1506788112
41. Bonomi, M., Camilloni, C.: Integrative structural and dynamical biology with PLUMED-ISDB. Bioinformatics. 33, 3999–4000 (2017). https://doi.org/10.1093/bioinformatics/btx529
42. Bonomi, M., Camilloni, C., Cavalli, A., Vendruscolo, M.: Metainference: A Bayesian inference method for heterogeneous systems. Sci Adv. 2, e1501177 (2016). https://doi.org/10.1126/sciadv.1501177
43. Bonomi, M., Branduardi, D., Bussi, G., Camilloni, C., Provasi, D., Raiteri, P., Donadio, D., Marinelli, F., Pietrucci, F., Broglia, R.A., Parrinello, M.: PLUMED: A portable plugin for free-energy calculations with molecular dynamics. Computer Physics Communications. 180, 1961–1972 (2009). https://doi.org/10.1016/j.cpc.2009.05.011
44. Tribello, G.A., Bonomi, M., Branduardi, D., Camilloni, C., Bussi, G.: PLUMED 2: New feathers for an old bird. Computer Physics Communications. 185, 604–613 (2014). https://doi.org/10.1016/j.cpc.2013.09.018
45. Różycki, B., Kim, Y.C., Hummer, G.: SAXS ensemble refinement of ESCRT-III CHMP3 conformational transitions. Structure. 19, 109–116 (2011). https://doi.org/10.1016/j.str.2010.10.006
46. Hummer, G., Köfinger, J.: Bayesian ensemble refinement by replica simulations and reweighting. The Journal of Chemical Physics. 143, 243150 (2015). https://doi.org/10.1063/1.4937786
47. Basta, J., Rauchman, M.: The nucleosome remodeling and deacetylase complex in development and disease. Translational Research. 165, 36–47 (2015). https://doi.org/10.1016/j.trsl.2014.05.003
48. Basenko, E.Y., Kamei, M., Ji, L., Schmitz, R.J., Lewis, Z.A.: The LSH/DDM1 Homolog MUS-30 Is Required for Genome Stability, but Not for DNA Methylation in Neurospora crassa. PLoS Genet. 12, e1005790 (2016). https://doi.org/10.1371/journal.pgen.1005790
49. Choi, D.-H., Kwon, S.-H., Kim, J.-H., Bae, S.-H.: Saccharomyces cerevisiae Cmr1 protein preferentially binds to UV-damaged DNA in vitro. J Microbiol. 50, 112–118 (2012). https://doi.org/10.1007/s12275-012-1597-4
50. Gallina, I., Colding, C., Henriksen, P., Beli, P., Nakamura, K., Offman, J., Mathiasen, D.P., Silva, S., Hoffmann, E., Groth, A., Choudhary, C., Lisby, M.: Cmr1/WDR76 defines a nuclear genotoxic stress body linking genome integrity and protein quality control. Nat Commun. 6, 6533 (2015). https://doi.org/10.1038/ncomms7533

51. Gilmore, J.M., Sardiu, M.E., Groppe, B.D., Thornton, J.L., Liu, X., Dayebgadoh, G., Banks, C.A., Slaughter, B.D., Unruh, J.R., Workman, J.L., Florens, L., Washburn, M.P.: WDR76 Co-Localizes with Heterochromatin Related Proteins and Rapidly Responds to DNA Damage. PLOS ONE. 11, e0155492 (2016). https://doi.org/10.1371/journal.pone.0155492
52. Gilmore, J.M., Sardiu, M.E., Venkatesh, S., Stutzman, B., Peak, A., Seidel, C.W., Workman, J.L., Florens, L., Washburn, M.P.: Characterization of a Highly Conserved Histone Related Protein, Ydl156w, and Its Functional Associations Using Quantitative Proteomic Analyses *. Molecular & Cellular Proteomics. 11, (2012). https://doi.org/10.1074/mcp.M111.011544
53. Higa, L.A., Wu, M., Ye, T., Kobayashi, R., Sun, H., Zhang, H.: CUL4–DDB1 ubiquitin ligase interacts with multiple WD40-repeat proteins and regulates histone methylation. Nat Cell Biol. 8, 1277–1283 (2006). https://doi.org/10.1038/ncb1490
54. Jeong, W.-J., Park, J.-C., Kim, W.-S., Ro, E.J., Jeon, S.H., Lee, S.-K., Park, Y.N., Min, D.S., Choi, K.-Y.: WDR76 is a RAS binding protein that functions as a tumor suppressor via RAS degradation. Nat Commun. 10, 295 (2019). https://doi.org/10.1038/s41467-018-08230-6
55. Jiang, Y., Mao, C., Yang, R., Yan, B., Shi, Y., Liu, X., Lai, W., Liu, Y., Wang, X., Xiao, D., Zhou, H., Cheng, Y., Yu, F., Cao, Y., Liu, S., Yan, Q., Tao, Y.: EGLN1/c-Myc Induced Lymphoid-Specific Helicase Inhibits Ferroptosis through Lipid Metabolic Gene Expression Changes. Theranostics. 7, 3293–3305 (2017). https://doi.org/10.7150/thno.19988
56. Jones, J.W., Singh, P., Govind, C.K.: Recruitment of Saccharomyces cerevisiae Cmr1/Ydl156w to Coding Regions Promotes Transcription Genome Wide. PLoS ONE. 11, e0148897 (2016). https://doi.org/10.1371/journal.pone.0148897
57. Liu, X., Zhang, Y., Wen, Z., Hao, Y., Banks, C.A.S., Lange, J.J., Cesare, J., Bhattacharya, S., Slaughter, B.D., Unruh, J.R., Florens, L., Workman, J.L., Washburn, M.P.: Serial Capture Affinity Purification and Integrated Structural Modeling of the H3K4me3 Binding and DNA Damage Related WDR76:SPIN1 Complex. Biochemistry (2023)
58. Park, J.-C., Jeong, W.-J., Seo, S.H., Choi, K.-Y.: WDR76 mediates obesity and hepatic steatosis via HRas destabilization. Sci Rep. 9, 19676 (2019). https://doi.org/10.1038/s41598-019-56211-6
59. Ro, E.J., Cho, Y.-H., Jeong, W.-J., Park, J.-C., Min, D.S., Choi, K.-Y.: WDR76 degrades RAS and suppresses cancer stem cell activation in colorectal cancer. Cell Commun Signal. 17, 88 (2019). https://doi.org/10.1186/s12964-019-0403-x
60. Tamayo, A.G., Duong, H.A., Robles, M.S., Mann, M., Weitz, C.J.: Histone monoubiquitination by Clock–Bmal1 complex marks Per1 and Per2 genes for circadian feedback. Nat Struct Mol Biol. 22, 759–766 (2015). https://doi.org/10.1038/nsmb.3076
61. Tkach, J.M., Yimit, A., Lee, A.Y., Riffle, M., Costanzo, M., Jaschob, D., Hendry, J.A., Ou, J., Moffat, J., Boone, C., Davis, T.N., Nislow, C., Brown, G.W.: Dissecting DNA damage response pathways by analysing protein localization and abundance changes during DNA replication stress. Nat Cell Biol. 14, 966–976 (2012). https://doi.org/10.1038/ncb2549
62. Weger, B.D., Sahinbas, M., Otto, G.W., Mracek, P., Armant, O., Dolle, D., Lahiri, K., Vallone, D., Ettwiller, L., Geisler, R., Foulkes, N.S., Dickmeis, T.: The Light Responsive Transcriptome of the Zebrafish: Function and Regulation. PLoS ONE. 6, e17080 (2011). https://doi.org/10.1371/journal.pone.0017080
63. Fang, L., Yu, G., Yu, W., Chen, G., Ye, B.: The correlation of WDR76 expression with survival outcomes and immune infiltrates in lung adenocarcinoma. PeerJ. 9, e12277 (2021). https://doi.org/10.7717/peerj.12277
64. Hu, Y., Tan, X., Zhang, L., Zhu, X., Wang, X.: WDR76 regulates 5-fluorouracil sensitivity in colon cancer via HRAS. Discov Onc. 14, 45 (2023). https://doi.org/10.1007/s12672-023-00656-9

65. Kwon, M., Oh, T., Jang, M., Kim, G.-H., Kim, J.-H., Ryu, H.W., Oh, S.-R., Jang, J.-H., Ahn, J.S., Ko, S.-K.: Kurarinone induced p53-independent G0/G1 cell cycle arrest by degradation of K-RAS via WDR76 in human colorectal cancer cells. European Journal of Pharmacology. 923, 174938 (2022). https://doi.org/10.1016/j.ejphar.2022.174938
66. Russel, D., Lasker, K., Webb, B., Velázquez-Muriel, J., Tjioe, E., Schneidman-Duhovny, D., Peterson, B., Sali, A.: Putting the Pieces Together: Integrative Modeling Platform Software for Structure Determination of Macromolecular Assemblies. PLoS Biol. 10, e1001244 (2012). https://doi.org/10.1371/journal.pbio.1001244
67. Padavattan, S., Thiruselvam, V., Shinagawa, T., Hasegawa, K., Kumasaka, T., Ishii, S., Kumarevel, T.: Structural analyses of the nucleosome complexes with human testis-specific histone variants, hTh2a and hTh2b. Biophysical Chemistry. 221, 41–48 (2017). https://doi.org/10.1016/j.bpc.2016.11.013
68. Jumper, J., Evans, R., Pritzel, A., Green, T., Figurnov, M., Ronneberger, O., Tunyasuvunakool, K., Bates, R., Žídek, A., Potapenko, A., Bridgland, A., Meyer, C., Kohl, S.A.A., Ballard, A.J., Cowie, A., Romera-Paredes, B., Nikolov, S., Jain, R., Adler, J., Back, T., Petersen, S., Reiman, D., Clancy, E., Zielinski, M., Steinegger, M., Pacholska, M., Berghammer, T., Bodenstein, S., Silver, D., Vinyals, O., Senior, A.W., Kavukcuoglu, K., Kohli, P., Hassabis, D.: Highly accurate protein structure prediction with AlphaFold. Nature. 596, 583–589 (2021). https://doi.org/10.1038/s41586-021-03819-2
69. Jeganathan, A., Leong, V., Zhao, L., Huen, J., Nano, N., Houry, W.A., Ortega, J.: Yeast Rvb1 and Rvb2 Proteins Oligomerize As a Conformationally Variable Dodecamer with Low Frequency. Journal of Molecular Biology. 427, 1875–1886 (2015). https://doi.org/10.1016/j.jmb.2015.01.010
70. Nano, N., Houry, W.A.: Chaperone-like activity of the AAA+ proteins Rvb1 and Rvb2 in the assembly of various complexes. Philosophical Transactions of the Royal Society B: Biological Sciences. 368, 20110399 (2013). https://doi.org/10.1098/rstb.2011.0399
71. Zhou, C.Y., Stoddard, C.I., Johnston, J.B., Trnka, M.J., Echeverria, I., Palovcak, E., Sali, A., Burlingame, A.L., Cheng, Y., Narlikar, G.J.: Regulation of Rvb1/Rvb2 by a Domain within the INO80 Chromatin Remodeling Complex Implicates the Yeast Rvbs as Protein Assembly Chaperones. Cell Reports. 19, 2033–2044 (2017). https://doi.org/10.1016/j.celrep.2017.05.029
72. Kim, M.: Beta conformation of polyglutamine track revealed by a crystal structure of Huntingtin N-terminal region with insertion of three histidine residues. Prion. 7, 221–228 (2013). https://doi.org/10.4161/pri.23807
73. Lakomek, K., Stoehr, G., Tosi, A., Schmailzl, M., Hopfner, K.-P.: Structural Basis for Dodecameric Assembly States and Conformational Plasticity of the Full-Length AAA+ ATPases Rvb1·Rvb2. Structure. 23, 483–495 (2015). https://doi.org/10.1016/j.str.2014.12.015
74. Lee, B.-G., Merkel, F., Allegretti, M., Hassler, M., Cawood, C., Lecomte, L., O'Reilly, F.J., Sinn, L.R., Gutierrez-Escribano, P., Kschonsak, M., Bravo, S., Nakane, T., Rappsilber, J., Aragon, L., Beck, M., Löwe, J., Haering, C.H.: Cryo-EM structures of holo condensin reveal a subunit flip-flop mechanism. Nat Struct Mol Biol. 27, 743–751 (2020). https://doi.org/10.1038/s41594-020-0457-x
75. Uhlmann, F.: SMC complexes: from DNA to chromosomes. Nat Rev Mol Cell Biol. 17, 399–412 (2016). https://doi.org/10.1038/nrm.2016.30
76. Aragón, L.: The Smc5/6 Complex: New and Old Functions of the Enigmatic Long-Distance Relative. Annu. Rev. Genet. 52, 89–107 (2018). https://doi.org/10.1146/annurev-genet-120417-031353
77. Kegel, A., Sjögren, C.: The Smc5/6 Complex: More Than Repair? Cold Spring Harb Symp Quant Biol. 75, 179–187 (2010). https://doi.org/10.1101/sqb.2010.75.047

78. Palecek, J.J.: SMC5/6: Multifunctional Player in Replication. Genes. 10, 7 (2019). https://doi.org/10.3390/genes10010007
79. Andrews, E.A., Palecek, J., Sergeant, J., Taylor, E., Lehmann, A.R., Watts, F.Z.: Nse2, a Component of the Smc5-6 Complex, Is a SUMO Ligase Required for the Response to DNA Damage. Molecular and Cellular Biology. 25, 185–196 (2005). https://doi.org/10.1128/MCB.25.1.185-196.2005
80. Potts, P.R., Yu, H.: Human MMS21/NSE2 Is a SUMO Ligase Required for DNA Repair. Molecular and Cellular Biology. 25, 7021–7032 (2005). https://doi.org/10.1128/MCB.25.16.7021-7032.2005
81. Solé-Soler, R., Torres-Rosell, J.: Smc5/6, an atypical SMC complex with two RING-type subunits. Biochemical Society Transactions. 48, 2159–2171 (2020). https://doi.org/10.1042/BST20200389
82. Zhao, X., Blobel, G.: A SUMO ligase is part of a nuclear multiprotein complex that affects DNA repair and chromosomal organization. Proceedings of the National Academy of Sciences. 102, 4777–4782 (2005). https://doi.org/10.1073/pnas.0500537102
83. Yu, Y., Li, S., Ser, Z., Sanyal, T., Choi, K., Wan, B., Kuang, H., Sali, A., Kentsis, A., Patel, D.J., Zhao, X.: Integrative analysis reveals unique structural and functional features of the Smc5/6 complex. Proceedings of the National Academy of Sciences. 118, e2026844118 (2021). https://doi.org/10.1073/pnas.2026844118
84. Basta, J., Rauchman, M.: The Nucleosome Remodeling and Deacetylase Complex in Development and Disease. In: Translating Epigenetics to the Clinic. pp. 37–72. Elsevier (2017)
85. Denslow, S.A., Wade, P.A.: The human Mi-2/NuRD complex and gene regulation. Oncogene. 26, 5433–5438 (2007). https://doi.org/10.1038/sj.onc.1210611
86. Toh, Y., Nicolson, G.L.: The role of the MTA family and their encoded proteins in human cancers: molecular functions and clinical implications. Clin Exp Metastasis. 26, 215–227 (2009). https://doi.org/10.1007/s10585-008-9233-8
87. Alqarni, S.S.M., Murthy, A., Zhang, W., Przewloka, M.R., Silva, A.P.G., Watson, A.A., Lejon, S., Pei, X.Y., Smits, A.H., Kloet, S.L., Wang, H., Shepherd, N.E., Stokes, P.H., Blobel, G.A., Vermeulen, M., Glover, D.M., Mackay, J.P., Laue, E.D.: Insight into the Architecture of the NuRD Complex. Journal of Biological Chemistry. 289, 21844–21855 (2014). https://doi.org/10.1074/jbc.M114.558940
88. Connelly, J.J., Yuan, P., Hsu, H.-C., Li, Z., Xu, R.-M., Sternglanz, R.: Structure and Function of the *Saccharomyces cerevisiae* Sir3 BAH Domain. Molecular and Cellular Biology. 26, 3256–3265 (2006). https://doi.org/10.1128/MCB.26.8.3256-3265.2006
89. Cramer, J.M., Scarsdale, J.N., Walavalkar, N.M., Buchwald, W.A., Ginder, G.D., Williams, D.C.: Probing the Dynamic Distribution of Bound States for Methylcytosine-binding Domains on DNA. Journal of Biological Chemistry. 289, 1294–1302 (2014). https://doi.org/10.1074/jbc.M113.512236
90. Gnanapragasam, M.N., Scarsdale, J.N., Amaya, M.L., Webb, H.D., Desai, M.A., Walavalkar, N.M., Wang, S.Z., Zu Zhu, S., Ginder, G.D., Williams, D.C.: p66α–MBD2 coiled-coil interaction and recruitment of Mi-2 are critical for globin gene silencing by the MBD2–NuRD complex. Proc. Natl. Acad. Sci. U.S.A. 108, 7487–7492 (2011). https://doi.org/10.1073/pnas.1015341108
91. Low, J.K.K., Silva, A.P.G., Sharifi Tabar, M., Torrado, M., Webb, S.R., Parker, B.L., Sana, M., Smits, C., Schmidberger, J.W., Brillault, L., Jackman, M.J., Williams, D.C., Blobel, G.A., Hake, S.B., Shepherd, N.E., Landsberg, M.J., Mackay, J.P.: The Nucleosome Remodeling and Deacetylase Complex Has an Asymmetric, Dynamic, and Modular Architecture. Cell Reports. 33, 108450 (2020). https://doi.org/10.1016/j.celrep.2020.108450
92. Millard, C.J., Varma, N., Saleh, A., Morris, K., Watson, P.J., Bottrill, A.R., Fairall, L., Smith, C.J., Schwabe, J.W.: The structure of the core NuRD repression complex provides insights into its interaction with chromatin. eLife. 5, e13941 (2016). https://doi.org/10.7554/eLife.13941
93. Millard, C.J., Watson, P.J., Celardo, I., Gordiyenko, Y., Cowley, S.M., Robinson, C.V., Fairall, L., Schwabe, J.W.R.: Class I HDACs Share a Common Mechanism of Regulation by Inositol Phosphates. Molecular Cell. 51, 57–67 (2013). https://doi.org/10.1016/j.molcel.2013.05.020
94. Arvindekar, S., Jackman, M.J., Low, J.K.K., Landsberg, M.J., Mackay, J.P., Viswanath, S.: Molecular architecture of nucleosome remodeling and deacetylase sub-complexes by integrative structure determination. Protein Science. 31, e4387 (2022). https://doi.org/10.1002/pro.4387
95. Desai, M.A., Webb, H.D., Sinanan, L.M., Scarsdale, J.N., Walavalkar, N.M., Ginder, G.D., Williams, D.C.: An intrinsically disordered region of methyl-CpG binding domain protein 2 (MBD2) recruits the histone deacetylase core of the NuRD complex. Nucleic Acids Research. 43, 3100–3113 (2015). https://doi.org/10.1093/nar/gkv168
96. Millard, C.J., Fairall, L., Ragan, T.J., Savva, C.G., Schwabe, J.W.R.: The topology of chromatin-binding domains in the NuRD deacetylase complex. Nucleic Acids Research. 48, 12972–12982 (2020). https://doi.org/10.1093/nar/gkaa1121
97. Pflum, M.K.H., Tong, J.K., Lane, W.S., Schreiber, S.L.: Histone Deacetylase 1 Phosphorylation Promotes Enzymatic Activity and Complex Formation. Journal of Biological Chemistry. 276, 47733–47741 (2001). https://doi.org/10.1074/jbc.M105590200
98. Zhang, Y., Ng, H.-H., Erdjument-Bromage, H., Tempst, P., Bird, A., Reinberg, D.: Analysis of the NuRD subunits reveals a histone deacetylase core complex and a connection with DNA methylation. Genes Dev. 13, 1924–1935 (1999)
99. Kaji, K., Caballero, I.M., MacLeod, R., Nichols, J., Wilson, V.A., Hendrich, B.: The NuRD component Mbd3 is required for pluripotency of embryonic stem cells. Nat Cell Biol. 8, 285–292 (2006). https://doi.org/10.1038/ncb1372
100. Liu, K., Lei, M., Wu, Z., Gan, B., Cheng, H., Li, Y., Min, J.: Structural analyses reveal that MBD3 is a methylated CG binder. The FEBS Journal. 286, 3240–3254 (2019). https://doi.org/10.1111/febs.14850
101. Chen, Z., Shiozaki, M., Haas, K.M., Zhao, S., Guo, C., Polacco, B.J., Yu, Z., Krogan, N.J., Kaake, R.M., Vale, R.D., Agard, D.A.: De novo protein identification in mammalian sperm using high-resolution in situ cryo-electron tomography, https://www.biorxiv.org/content/10.1101/2022.09.28.510016v1, (2022)
102. Leung, M.R., Roelofs, M.C., Chiozzi, R.Z., Hevler, J.F., Heck, A.J.R., Zeev-Ben-Mordehai, T.: Unraveling the intricate microtubule inner protein networks that reinforce mammalian sperm flagella, https://www.biorxiv.org/content/10.1101/2022.09.29.510157v1, (2022)
103. Hesketh, S.J., Mukhopadhyay, A.G., Nakamura, D., Toropova, K., Roberts, A.J.: IFT-A structure reveals carriages for membrane protein transport into cilia. Cell. 185, 4971-4985.e16 (2022). https://doi.org/10.1016/j.cell.2022.11.010
104. Gui, M., Wang, X., Dutcher, S.K., Brown, A., Zhang, R.: Ciliary central apparatus structure reveals mechanisms of microtubule patterning. Nat Struct Mol Biol. 29, 483–492 (2022). https://doi.org/10.1038/s41594-022-00770-2
105. Kowalczyk, A.P., Green, K.J.: Structure, Function, and Regulation of Desmosomes. In: Progress in Molecular Biology and Translational Science. pp. 95–118. Elsevier (2013)
106. Pasani, S., Menon, K.S., Viswanath, S.: The molecular architecture of the desmosomal outer dense plaque by integrative structural modeling. Biophysics (2023)
107. Sikora, M., Ermel, U.H., Seybold, A., Kunz, M., Calloni, G., Reitz, J., Vabulas, R.M., Hummer, G., Frangakis, A.S.: Desmosome architecture derived from molecular dynamics simulations and cryo-electron tomography. Proc. Natl. Acad. Sci. U.S.A. 117, 27132–27140 (2020). https://doi.org/10.1073/pnas.2004563117

108. Fontana, P., Dong, Y., Pi, X., Tong, A.B., Hecksel, C.W., Wang, L., Fu, T.-M., Bustamante, C., Wu, H.: Structure of cytoplasmic ring of nuclear pore complex by integrative cryo-EM and AlphaFold. Science. 376, eabm9326 (2022). https://doi.org/10.1126/science.abm9326
109. Zhu, X., Huang, G., Zeng, C., Zhan, X., Liang, K., Xu, Q., Zhao, Y., Wang, P., Wang, Q., Zhou, Q., Tao, Q., Liu, M., Lei, J., Yan, C., Shi, Y.: Structure of the cytoplasmic ring of the *Xenopus laevis* nuclear pore complex. Science. 376, eabl8280 (2022). https://doi.org/10.1126/science.abl8280
110. Bley, C.J., Nie, S., Mobbs, G.W., Petrovic, S., Gres, A.T., Liu, X., Mukherjee, S., Harvey, S., Huber, F.M., Lin, D.H., Brown, B., Tang, A.W., Rundlet, E.J., Correia, A.R., Chen, S., Regmi, S.G., Stevens, T.A., Jette, C.A., Dasso, M., Patke, A., Palazzo, A.F., Kossiakoff, A.A., Hoelz, A.: Architecture of the cytoplasmic face of the nuclear pore. Science. 376, eabm9129 (2022). https://doi.org/10.1126/science.abm9129
111. Petrovic, S., Samanta, D., Perriches, T., Bley, C.J., Thierbach, K., Brown, B., Nie, S., Mobbs, G.W., Stevens, T.A., Liu, X., Tomaleri, G.P., Schaus, L., Hoelz, A.: Architecture of the linker-scaffold in the nuclear pore. Science. 376, eabm9798 (2022). https://doi.org/10.1126/science.abm9798
112. Akey, C.W., Echeverria, I., Ouch, C., Nudelman, I., Shi, Y., Wang, J., Chait, B.T., Sali, A., Fernandez-Martinez, J., Rout, M.P.: Implications of a multiscale structure of the yeast nuclear pore complex. Molecular Cell. 83, 3283-3302.e5 (2023). https://doi.org/10.1016/j.molcel.2023.08.025
113. Akey, C.W., Singh, D., Ouch, C., Echeverria, I., Nudelman, I., Varberg, J.M., Yu, Z., Fang, F., Shi, Y., Wang, J., Salzberg, D., Song, K., Xu, C., Gumbart, J.C., Suslov, S., Unruh, J., Jaspersen, S.L., Chait, B.T., Sali, A., Fernandez-Martinez, J., Ludtke, S.J., Villa, E., Rout, M.P.: Comprehensive structure and functional adaptations of the yeast nuclear pore complex. Cell. 185, 361-378.e25 (2022). https://doi.org/10.1016/j.cell.2021.12.015
114. Mosalaganti, S., Obarska-Kosinska, A., Siggel, M., Taniguchi, R., Turoňová, B., Zimmerli, C.E., Buczak, K., Schmidt, F.H., Margiotta, E., Mackmull, M.-T., Hagen, W.J.H., Hummer, G., Kosinski, J., Beck, M.: AI-based structure prediction empowers integrative structural analysis of human nuclear pores. Science. 376, eabm9506 (2022). https://doi.org/10.1126/science.abm9506
115. Otsuka, S., Tempkin, J.O.B., Zhang, W., Politi, A.Z., Rybina, A., Hossain, M.J., Kueblbeck, M., Callegari, A., Koch, B., Morero, N.R., Sali, A., Ellenberg, J.: A quantitative map of nuclear pore assembly reveals two distinct mechanisms. Nature. 613, 575–581 (2023). https://doi.org/10.1038/s41586-022-05528-w
116. Flacht, L., Lunelli, M., Kaszuba, K., Chen, Z.A., Reilly, F.J.O., Rappsilber, J., Kosinski, J., Kolbe, M.: Integrative structural analysis of the type III secretion system needle complex from *Shigella flexneri*. Protein Science. 32, e4595 (2023). https://doi.org/10.1002/pro.4595
117. Cohen, T., Halfon, M., Schneidman-Duhovny, D.: NanoNet: Rapid and accurate end-to-end nanobody modeling by deep learning. Front Immunol. 13, 958584 (2022). https://doi.org/10.3389/fimmu.2022.958584
118. Xiang, Y., Sang, Z., Bitton, L., Xu, J., Liu, Y., Schneidman-Duhovny, D., Shi, Y.: Integrative proteomics identifies thousands of distinct, multi-epitope, and high-affinity nanobodies. Cell Syst. 12, 220-234.e9 (2021). https://doi.org/10.1016/j.cels.2021.01.003
119. Muyldermans, S.: Applications of Nanobodies. Annu Rev Anim Biosci. 9, 401–421 (2021). https://doi.org/10.1146/annurev-animal-021419-083831
120. Xiang, Y., Nambulli, S., Xiao, Z., Liu, H., Sang, Z., Duprex, W.P., Schneidman-Duhovny, D., Zhang, C., Shi, Y.: Versatile and multivalent nanobodies efficiently neutralize SARS-CoV-2. Science. 370, 1479–1484 (2020). https://doi.org/10.1126/science.abe4747
121. Giulini, M., Schneider, C., Cutting, D., Desai, N., Deane, C.M., Bonvin, A.M.J.J.: Towards the accurate modelling of antibody-antigen complexes from sequence using machine learning and information-driven docking, https://www.biorxiv.org/content/10.1101/2023.11.17.567543v1, (2023)
122. Abanades, B., Wong, W.K., Boyles, F., Georges, G., Bujotzek, A., Deane, C.M.: ImmuneBuilder: Deep-Learning models for predicting the structures of immune proteins. Commun Biol. 6, 1–8 (2023). https://doi.org/10.1038/s42003-023-04927-7
123. Evans, R., O'Neill, M., Pritzel, A., Antropova, N., Senior, A., Green, T., Žídek, A., Bates, R., Blackwell, S., Yim, J., Ronneberger, O., Bodenstein, S., Zielinski, M., Bridgland, A., Potapenko, A., Cowie, A., Tunyasuvunakool, K., Jain, R., Clancy, E., Kohli, P., Jumper, J., Hassabis, D.: Protein complex prediction with AlphaFold-Multimer, https://www.biorxiv.org/content/10.1101/2021.10.04.463034v2, (2022)
124. Mani, R., St.Onge, R.P., Hartman, J.L., Giaever, G., Roth, F.P.: Defining genetic interaction. Proc. Natl. Acad. Sci. U.S.A. 105, 3461–3466 (2008). https://doi.org/10.1073/pnas.0712255105
125. Braberg, H., Echeverria, I., Bohn, S., Cimermancic, P., Shiver, A., Alexander, R., Xu, J., Shales, M., Dronamraju, R., Jiang, S., Dwivedi, G., Bogdanoff, D., Chaung, K.K., Hüttenhain, R., Wang, S., Mavor, D., Pellarin, R., Schneidman, D., Bader, J.S., Fraser, J.S., Morris, J., Haber, J.E., Strahl, B.D., Gross, C.A., Dai, J., Boeke, J.D., Sali, A., Krogan, N.J.: Genetic interaction mapping informs integrative structure determination of protein complexes. Science. 370, eaaz4910 (2020). https://doi.org/10.1126/science.aaz4910
126. Echeverria, I., Braberg, H., Krogan, N.J., Sali, A.: Integrative structure determination of histones H3 and H4 using genetic interactions. The FEBS Journal. 290, 2565–2575 (2023). https://doi.org/10.1111/febs.16435
127. Kastritis, P.L., O'Reilly, F.J., Bock, T., Li, Y., Rogon, M.Z., Buczak, K., Romanov, N., Betts, M.J., Bui, K.H., Hagen, W.J., Hennrich, M.L., Mackmull, M., Rappsilber, J., Russell, R.B., Bork, P., Beck, M., Gavin, A.: Capturing protein communities by structural proteomics in a thermophilic eukaryote. Molecular Systems Biology. 13, 936 (2017). https://doi.org/10.15252/msb.20167412
128. Skalidis, I., Kyrilis, F.L., Tüting, C., Hamdi, F., Träger, T.K., Belapure, J., Hause, G., Fratini, M., O'Reilly, F.J., Heilmann, I., Rappsilber, J., Kastritis, P.L.: Structural analysis of an endogenous 4-megadalton succinyl-CoA-generating metabolon. Commun Biol. 6, 552 (2023). https://doi.org/10.1038/s42003-023-04885-0
129. Kyrilis, F.L., Semchonok, D.A., Skalidis, I., Tüting, C., Hamdi, F., O'Reilly, F.J., Rappsilber, J., Kastritis, P.L.: Integrative structure of a 10-megadalton eukaryotic pyruvate dehydrogenase complex from native cell extracts. Cell Reports. 34, 108727 (2021). https://doi.org/10.1016/j.celrep.2021.108727
130. Tüting, C., Kyrilis, F.L., Müller, J., Sorokina, M., Skalidis, I., Hamdi, F., Sadian, Y., Kastritis, P.L.: Cryo-EM snapshots of a native lysate provide structural insights into a metabolon-embedded transacetylase reaction. Nat Commun. 12, 6933 (2021). https://doi.org/10.1038/s41467-021-27287-4
131. Uversky, V.N.: Intrinsically Disordered Proteins and Their "Mysterious" (Meta)Physics. Front. Phys. 7, 10 (2019). https://doi.org/10.3389/fphy.2019.00010
132. Hoff, S.E., Thomasen, F.E., Lindorff-Larsen, K., Bonomi, M.: Accurate model and ensemble refinement using cryo-electron microscopy maps and Bayesian inference. Bioinformatics (2023)
133. Nogales, E.: The development of cryo-EM into a mainstream structural biology technique. Nat Methods. 13, 24–27 (2016). https://doi.org/10.1038/nmeth.3694
134. Wang, X., Alnabati, E., Aderinwale, T.W., Venkata Subramaniya, S.R.M., Terashi, G., Kihara, D.: Emap2sec+: Detecting Protein and DNA/RNA Structures in Cryo-EM Maps of Intermediate Resolution Using Deep Learning. Biophysics (2020)

135. Scheres, S.H.W.: Chapter Six - Processing of Structurally Heterogeneous Cryo-EM Data in RELION. In: Crowther, R.A. (ed.) Methods in Enzymology. pp. 125–157. Academic Press (2016)
136. Schwab, J., Kimanius, D., Burt, A., Dendooven, T., Scheres, S.H.W.: DynaMight: estimating molecular motions with improved reconstruction from cryo-EM images. Biophysics (2023)
137. Bai, X., Rajendra, E., Yang, G., Shi, Y., Scheres, S.H.: Sampling the conformational space of the catalytic subunit of human γ-secretase. eLife. 4, e11182 (2015). https://doi.org/10.7554/eLife.11182
138. Chen, M., Ludtke, S.J.: Deep learning-based mixed-dimensional Gaussian mixture model for characterizing variability in cryo-EM. Nat Methods. 18, 930–936 (2021). https://doi.org/10.1038/s41592-021-01220-5
139. Kimanius, D., Jamali, K., Scheres, S.H.W.: Sparse Fourier Backpropagation in Cryo-EM Reconstruction.
140. Nakane, T., Kimanius, D., Lindahl, E., Scheres, S.H.: Characterisation of molecular motions in cryo-EM single-particle data by multi-body refinement in RELION. eLife. 7, e36861 (2018). https://doi.org/10.7554/eLife.36861
141. Zhong, E.D., Bepler, T., Berger, B., Davis, J.H.: CryoDRGN: reconstruction of heterogeneous cryo-EM structures using neural networks. Nat Methods. 18, 176–185 (2021). https://doi.org/10.1038/s41592-020-01049-4
142. Zhou, Q., Huang, X., Sun, S., Li, X., Wang, H.-W., Sui, S.-F.: Cryo-EM structure of SNAP-SNARE assembly in 20S particle. Cell Res. 25, 551–560 (2015). https://doi.org/10.1038/cr.2015.47
143. Zhong, E.D., Lerer, A., Davis, J.H., Berger, B.: CryoDRGN2: Ab initio neural reconstruction of 3D protein structures from real cryo-EM images. In: 2021 IEEE/CVF International Conference on Computer Vision (ICCV). pp. 4046–4055. IEEE, Montreal, QC, Canada (2021)
144. Zhong, E.D., Bepler, T., Davis, J.H., Berger, B.: Reconstructing continuous distributions of 3D protein structure from cryo-EM images, http://arxiv.org/abs/1909.05215, (2020)
145. IMP, the Integrative Modeling Platform, https://integrativemodeling.org/
146. Bartolec, T.K., Vázquez-Campos, X., Norman, A., Luong, C., Johnson, M., Payne, R.J., Wilkins, M.R., Mackay, J.P., Low, J.K.K.: Cross-linking mass spectrometry discovers, evaluates, and corroborates structures and protein–protein interactions in the human cell. Proceedings of the National Academy of Sciences. 120, e2219418120 (2023). https://doi.org/10.1073/pnas.2219418120
147. Sinz, A., Arlt, C., Chorev, D., Sharon, M.: Chemical cross-linking and native mass spectrometry: A fruitful combination for structural biology. Protein Science. 24, 1193–1209 (2015). https://doi.org/10.1002/pro.2696
148. Ziemianowicz, D.S., Saltzberg, D., Pells, T., Crowder, D.A., Schräder, C., Hepburn, M., Sali, A., Schriemer, D.C.: IMProv: A Resource for Cross-link-Driven Structure Modeling that Accommodates Protein Dynamics. Molecular & Cellular Proteomics. 20, 100139 (2021). https://doi.org/10.1016/j.mcpro.2021.100139
149. Bullock, J.M.A., Sen, N., Thalassinos, K., Topf, M.: Modeling Protein Complexes Using Restraints from Crosslinking Mass Spectrometry. Structure. 26, 1015-1024.e2 (2018). https://doi.org/10.1016/j.str.2018.04.016
150. Sinnott, M., Malhotra, S., Madhusudhan, M.S., Thalassinos, K., Topf, M.: Combining Information from Crosslinks and Monolinks in the Modeling of Protein Structures. Structure. 28, 1061-1070.e3 (2020). https://doi.org/10.1016/j.str.2020.05.012
151. Cohen, S., Schneidman-Duhovny, D.: A deep learning model for predicting optimal distance range in crosslinking mass spectrometry data. Proteomics. 23, 2200341 (2023). https://doi.org/10.1002/pmic.202200341

152. Viswanath, S., Sali, A.: Optimizing model representation for integrative structure determination of macromolecular assemblies. Proc. Natl. Acad. Sci. U.S.A. 116, 540–545 (2019). https://doi.org/10.1073/pnas.1814649116
153. Arvindekar, S., Pathak, A.S., Majila, K., Viswanath, S.: Optimizing representations for integrative structural modeling using Bayesian model selection. Bioinformatics (2023)
154. Bonomi, M., Heller, G.T., Camilloni, C., Vendruscolo, M.: Principles of protein structural ensemble determination. Current Opinion in Structural Biology. 42, 106–116 (2017). https://doi.org/10.1016/j.sbi.2016.12.004
155. Carstens, S., Nilges, M., Habeck, M.: Bayesian inference of chromatin structure ensembles from population-averaged contact data. Proc. Natl. Acad. Sci. U.S.A. 117, 7824–7830 (2020). https://doi.org/10.1073/pnas.1910364117
156. Ge, Y., Voelz, V.A.: Model Selection Using BICePs: A Bayesian Approach for Force Field Validation and Parameterization. J. Phys. Chem. B. 122, 5610–5622 (2018). https://doi.org/10.1021/acs.jpcb.7b11871
157. Voelz, V.A., Ge, Y., Raddi, R.M.: Reconciling Simulations and Experiments With BICePs: A Review. Front. Mol. Biosci. 8, 661520 (2021). https://doi.org/10.3389/fmolb.2021.661520
158. Bonomi, M., Camilloni, C., Vendruscolo, M.: Metadynamic metainference: Enhanced sampling of the metainference ensemble using metadynamics. Sci Rep. 6, 31232 (2016). https://doi.org/10.1038/srep31232
159. Bottaro, S., Bengtsen, T., Lindorff-Larsen, K.: Integrating Molecular Simulation and Experimental Data: A Bayesian/Maximum Entropy Reweighting Approach. In: Gáspári, Z. (ed.) Structural Bioinformatics. pp. 219–240. Springer US, New York, NY (2020)
160. Lincoff, J., Haghighatlari, M., Krzeminski, M., Teixeira, J.M.C., Gomes, G.-N.W., Gradinaru, C.C., Forman-Kay, J.D., Head-Gordon, T.: Extended experimental inferential structure determination method in determining the structural ensembles of disordered protein states. Commun Chem. 3, 1–12 (2020). https://doi.org/10.1038/s42004-020-0323-0
161. Pasani, S., Viswanath, S.: A Framework for Stochastic Optimization of Parameters for Integrative Modeling of Macromolecular Assemblies. Life. 11, 1183 (2021). https://doi.org/10.3390/life11111183
162. Ullanat, V., Kasukurthi, N., Viswanath, S.: PrISM: precision for integrative structural models. Bioinformatics. 38, 3837–3839 (2022). https://doi.org/10.1093/bioinformatics/btac400
163. Kühlbrandt, W.: Biochemistry. The resolution revolution. Science. 343, 1443–1444 (2014). https://doi.org/10.1126/science.1251652
164. Baek, M., DiMaio, F., Anishchenko, I., Dauparas, J., Ovchinnikov, S., Lee, G.R., Wang, J., Cong, Q., Kinch, L.N., Schaeffer, R.D., Millán, C., Park, H., Adams, C., Glassman, C.R., DeGiovanni, A., Pereira, J.H., Rodrigues, A.V., van Dijk, A.A., Ebrecht, A.C., Opperman, D.J., Sagmeister, T., Buhlheller, C., Pavkov-Keller, T., Rathinaswamy, M.K., Dalwadi, U., Yip, C.K., Burke, J.E., Garcia, K.C., Grishin, N.V., Adams, P.D., Read, R.J., Baker, D.: Accurate prediction of protein structures and interactions using a three-track neural network. Science. 373, 871–876 (2021). https://doi.org/10.1126/science.abj8754
165. Terwilliger, T.C., Afonine, P.V., Liebschner, D., Croll, T.I., McCoy, A.J., Oeffner, R.D., Williams, C.J., Poon, B.K., Richardson, J.S., Read, R.J., Adams, P.D.: Accelerating crystal structure determination with iterative AlphaFold prediction. Acta Cryst D. 79, 234–244 (2023). https://doi.org/10.1107/S205979832300102X
166. Gao, M., Nakajima An, D., Parks, J.M., Skolnick, J.: AF2Complex predicts direct physical interactions in multimeric proteins with deep learning. Nat Commun. 13, 1744 (2022). https://doi.org/10.1038/s41467-022-29394-2
167. McCafferty, C.L., Pennington, E.L., Papoulas, O., Taylor, D.W., Marcotte, E.M.: Does AlphaFold2 model proteins' intracellular conformations? An experimental test using cross-linking mass spectrometry of endogenous ciliary proteins. Commun Biol. 6, 421 (2023). https://doi.org/10.1038/s42003-023-04773-7

168. Humphreys, I.R., Pei, J., Baek, M., Krishnakumar, A., Anishchenko, I., Ovchinnikov, S., Zhang, J., Ness, T.J., Banjade, S., Bagde, S.R., Stancheva, V.G., Li, X.-H., Liu, K., Zheng, Z., Barrero, D.J., Roy, U., Kuper, J., Fernández, I.S., Szakal, B., Branzei, D., Rizo, J., Kisker, C., Greene, E.C., Biggins, S., Keeney, S., Miller, E.A., Fromme, J.C., Hendrickson, T.L., Cong, Q., Baker, D.: Computed structures of core eukaryotic protein complexes. Science. 374, eabm4805 (2021). https://doi.org/10.1126/science.abm4805
169. Burke, D.F., Bryant, P., Barrio-Hernandez, I., Memon, D., Shenoy, A., Zhu, W., Dunham, A.S., Albanese, P., Scheltema, R.A., Bruce, J.E., Leitner, A., Kundrotas, P., Beltrao, P., Elofsson, A.: Towards a structurally resolved human protein interaction network.
170. O'Reilly, F.J., Graziadei, A., Forbrig, C., Bremenkamp, R., Charles, K., Lenz, S., Elfmann, C., Fischer, L., Stülke, J., Rappsilber, J.: Protein complexes in cells by AI-assisted structural proteomics. Molecular Systems Biology. 19, e11544 (2023). https://doi.org/10.15252/msb.202311544
171. Stahl, K., Brock, O., Rappsilber, J.: Modelling protein complexes with crosslinking mass spectrometry and deep learning, https://www.biorxiv.org/content/10.1101/2023.06.07.544059v2, (2023)
172. Mirabello, C., Wallner, B., Nystedt, B., Azinas, S., Carroni, M.: Unmasking AlphaFold: integration of experiments and predictions in multimeric complexes, https://www.biorxiv.org/content/10.1101/2023.09.20.558579v3, (2023)
173. Terwilliger, T.C., Poon, B.K., Afonine, P.V., Schlicksup, C.J., Croll, T.I., Millán, C., Richardson, J.S., Read, R.J., Adams, P.D.: Improved AlphaFold modeling with implicit experimental information. Nat Methods. 19, 1376–1382 (2022). https://doi.org/10.1038/s41592-022-01645-6
174. Watson, J.L., Juergens, D., Bennett, N.R., Trippe, B.L., Yim, J., Eisenach, H.E., Ahern, W., Borst, A.J., Ragotte, R.J., Milles, L.F., Wicky, B.I.M., Hanikel, N., Pellock, S.J., Courbet, A., Sheffler, W., Wang, J., Venkatesh, P., Sappington, I., Torres, S.V., Lauko, A., De Bortoli, V., Mathieu, E., Ovchinnikov, S., Barzilay, R., Jaakkola, T.S., DiMaio, F., Baek, M., Baker, D.: De novo design of protein structure and function with RFdiffusion. Nature. 620, 1089–1100 (2023). https://doi.org/10.1038/s41586-023-06415-8
175. Baul, U., Chakraborty, D., Mugnai, M.L., Straub, J.E., Thirumalai, D.: Sequence Effects on Size, Shape, and Structural Heterogeneity in Intrinsically Disordered Proteins. J. Phys. Chem. B. 123, 3462–3474 (2019). https://doi.org/10.1021/acs.jpcb.9b02575
176. Zhang, O., Haghighatlari, M., Li, J., Liu, Z.H., Namini, A., Teixeira, J.M.C., Forman-Kay, J.D., Head-Gordon, T.: Learning to evolve structural ensembles of unfolded and disordered proteins using experimental solution data. J Chem Phys. 158, 174113 (2023). https://doi.org/10.1063/5.0141474
177. Lamm, L., Righetto, R.D., Wietrzynski, W., Pöge, M., Martinez-Sanchez, A., Peng, T., Engel, B.D.: MemBrain: A deep learning-aided pipeline for detection of membrane proteins in Cryo-electron tomograms. Computer Methods and Programs in Biomedicine. 224, 106990 (2022). https://doi.org/10.1016/j.cmpb.2022.106990
178. Gubins, I., Chaillet, M.L., van der Schot, G., Veltkamp, R.C., Förster, F., Hao, Y., Wan, X., Cui, X., Zhang, F., Moebel, E., Wang, X., Kihara, D., Zeng, X., Xu, M., Nguyen, N.P., White, T., Bunyak, F.: SHREC 2020: Classification in cryo-electron tomograms. Computers & Graphics. 91, 279–289 (2020). https://doi.org/10.1016/j.cag.2020.07.010
179. Pyle, E., Zanetti, G.: Current data processing strategies for cryo-electron tomography and subtomogram averaging. Biochemical Journal. 478, 1827–1845 (2021). https://doi.org/10.1042/BCJ20200715
180. Moebel, E., Martinez-Sanchez, A., Lamm, L., Righetto, R.D., Wietrzynski, W., Albert, S., Larivière, D., Fourmentin, E., Pfeffer, S., Ortiz, J., Baumeister, W., Peng, T., Engel, B.D.,

Kervrann, C.: Deep learning improves macromolecule identification in 3D cellular cryo-electron tomograms. Nat Methods. 18, 1386–1394 (2021). https://doi.org/10.1038/s41592-021-01275-4
181. Zeng, X., Kahng, A., Xue, L., Mahamid, J., Chang, Y.-W., Xu, M.: High-throughput cryo-ET structural pattern mining by unsupervised deep iterative subtomogram clustering. Proceedings of the National Academy of Sciences. 120, e2213149120 (2023). https://doi.org/10.1073/pnas.2213149120
182. Förster, F., Han, B.-G., Beck, M.: Chapter Eleven - Visual Proteomics. In: Jensen, G.J. (ed.) Methods in Enzymology. pp. 215–243. Academic Press (2010)
183. Kim, H.H.-S., Uddin, M.R., Xu, M., Chang, Y.-W.: Computational Methods Toward Unbiased Pattern Mining and Structure Determination in Cryo-Electron Tomography Data. Journal of Molecular Biology. 435, 168068 (2023). https://doi.org/10.1016/j.jmb.2023.168068
184. Gubins, I., Chaillet, M.L., Schot, G.V.D., Trueba, M.C., Veltkamp, R.C., Förster, F., Wang, X., Kihara, D., Moebel, E., Nguyen, N.P., White, T., Bunyak, F., Papoulias, G., Gerolymatos, S., Zacharaki, E.I., Moustakas, K., Zeng, X., Liu, S., Xu, M., Wang, Y., Chen, C., Cui, X., Zhang, F.: SHREC 2021: Classification in Cryo-electron Tomograms. Eurographics Workshop on 3D Object Retrieval. 13 pages (2021). https://doi.org/10.2312/3DOR.20211307
185. Gubins, I., Schot, G.V.D., Veltkamp, R.C., Förster, F., Du, X., Zeng, X., Zhu, Z., Chang, L., Xu, M., Moebel, E., Martinez-Sanchez, A., Kervrann, C., Lai, T.M., Han, X., Terashi, G., Kihara, D., Himes, B.A., Wan, X., Zhang, J., Gao, S., Hao, Y., Lv, Z., Wan, X., Yang, Z., Ding, Z., Cui, X., Zhang, F.: Classification in Cryo-Electron Tomograms. Eurographics Workshop on 3D Object Retrieval. 6 pages (2019). https://doi.org/10.2312/3DOR.20191061
186. Frangakis, A.S., Böhm, J., Förster, F., Nickell, S., Nicastro, D., Typke, D., Hegerl, R., Baumeister, W.: Identification of macromolecular complexes in cryoelectron tomograms of phantom cells. Proceedings of the National Academy of Sciences. 99, 14153–14158 (2002). https://doi.org/10.1073/pnas.172520299
187. Li, R., Yu, L., Zhou, B., Zeng, X., Wang, Z., Yang, X., Zhang, J., Gao, X., Jiang, R., Xu, M.: Few-shot learning for classification of novel macromolecular structures in cryo-electron tomograms. PLOS Computational Biology. 16, e1008227 (2020). https://doi.org/10.1371/journal.pcbi.1008227
188. Rice, G., Wagner, T., Stabrin, M., Sitsel, O., Prumbaum, D., Raunser, S.: TomoTwin: generalized 3D localization of macromolecules in cryo-electron tomograms with structural data mining. Nat Methods. 20, 871–880 (2023). https://doi.org/10.1038/s41592-023-01878-z
189. de Teresa-Trueba, I., Goetz, S.K., Mattausch, A., Stojanovska, F., Zimmerli, C.E., Toro-Nahuelpan, M., Cheng, D.W.C., Tollervey, F., Pape, C., Beck, M., Diz-Muñoz, A., Kreshuk, A., Mahamid, J., Zaugg, J.B.: Convolutional networks for supervised mining of molecular patterns within cellular context. Nat Methods. 20, 284–294 (2023). https://doi.org/10.1038/s41592-022-01746-2
190. Wagner, T., Merino, F., Stabrin, M., Moriya, T., Antoni, C., Apelbaum, A., Hagel, P., Sitsel, O., Raisch, T., Prumbaum, D., Quentin, D., Roderer, D., Tacke, S., Siebolds, B., Schubert, E., Shaikh, T.R., Lill, P., Gatsogiannis, C., Raunser, S.: SPHIRE-crYOLO is a fast and accurate fully automated particle picker for cryo-EM. Commun Biol. 2, 1–13 (2019). https://doi.org/10.1038/s42003-019-0437-z
191. Xu, M., Singla, J., Tocheva, E.I., Chang, Y.-W., Stevens, R.C., Jensen, G.J., Alber, F.: De Novo Structural Pattern Mining in Cellular Electron Cryotomograms. Structure. 27, 679-691.e14 (2019). https://doi.org/10.1016/j.str.2019.01.005

192. Yu, L., Li, R., Zeng, X., Wang, H., Jin, J., Ge, Y., Jiang, R., Xu, M.: Few shot domain adaptation for in situ macromolecule structural classification in cryoelectron tomograms. Bioinformatics. 37, 185–191 (2021). https://doi.org/10.1093/bioinformatics/btaa671
193. Johnson, G.T., Autin, L., Al-Alusi, M., Goodsell, D.S., Sanner, M.F., Olson, A.J.: cellPACK: a virtual mesoscope to model and visualize structural systems biology. Nat Methods. 12, 85–91 (2015). https://doi.org/10.1038/nmeth.3204
194. Maritan, M., Autin, L., Karr, J., Covert, M.W., Olson, A.J., Goodsell, D.S.: Building Structural Models of a Whole Mycoplasma Cell. Journal of Molecular Biology. 434, 167351 (2022). https://doi.org/10.1016/j.jmb.2021.167351
195. Luthey-Schulten, Z., Thornburg, Z.R., Gilbert, B.R.: Integrating cellular and molecular structures and dynamics into whole-cell models. Current Opinion in Structural Biology. 75, 102392 (2022). https://doi.org/10.1016/j.sbi.2022.102392
196. Thornburg, Z.R., Bianchi, D.M., Brier, T.A., Gilbert, B.R., Earnest, T.M., Melo, M.C.R., Safronova, N., Sáenz, J.P., Cook, A.T., Wise, K.S., Hutchison, C.A., Smith, H.O., Glass, J.I., Luthey-Schulten, Z.: Fundamental behaviors emerge from simulations of a living minimal cell. Cell. 185, 345-360.e28 (2022). https://doi.org/10.1016/j.cell.2021.12.025
197. Stevens, J.A., Grünewald, F., Van Tilburg, P.A.M., König, M., Gilbert, B.R., Brier, T.A., Thornburg, Z.R., Luthey-Schulten, Z., Marrink, S.J.: Molecular dynamics simulation of an entire cell. Front. Chem. 11, 1106495 (2023). https://doi.org/10.3389/fchem.2023.1106495
198. Raveh, B., Sun, L., White, K.L., Sanyal, T., Tempkin, J., Zheng, D., Bharath, K., Singla, J., Wang, C., Zhao, J., Li, A., Graham, N.A., Kesselman, C., Stevens, R.C., Sali, A.: Bayesian metamodeling of complex biological systems across varying representations. Proc. Natl. Acad. Sci. U.S.A. 118, e2104559118 (2021). https://doi.org/10.1073/pnas.2104559118
199. Lieberman, R., Mintz, R., Raveh, B.: Bayesian Metamodeling of pancreatic islet architecture and functional dynamics. Systems Biology (2021)
200. Berman, H.M., Adams, P.D., Bonvin, A.A., Burley, S.K., Carragher, B., Chiu, W., DiMaio, F., Ferrin, T.E., Gabanyi, M.J., Goddard, T.D., Griffin, P.R., Haas, J., Hanke, C.A., Hoch, J.C., Hummer, G., Kurisu, G., Lawson, C.L., Leitner, A., Markley, J.L., Meiler, J., Montelione, G.T., Phillips, G.N., Prisner, T., Rappsilber, J., Schriemer, D.C., Schwede, T., Seidel, C.A.M., Strutzenberg, T.S., Svergun, D.I., Tajkhorshid, E., Trewhella, J., Vallat, B., Velankar, S., Vuister, G.W., Webb, B., Westbrook, J.D., White, K.L., Sali, A.: Federating Structural Models and Data: Outcomes from A Workshop on Archiving Integrative Structures. Structure. 27, 1745–1759 (2019). https://doi.org/10.1016/j.str.2019.11.002